\documentclass[aps,prr,twocolumn,superscriptaddress,nofootinbib]{revtex4}
\usepackage{amsfonts}
\usepackage{amsmath}
\usepackage{mathtools}
\usepackage{longtable}
\usepackage{amssymb}
\usepackage{graphicx}
\usepackage{bm}
\usepackage{color}
\usepackage{mathrsfs}
\usepackage[colorlinks,bookmarks=true,citecolor=blue,linkcolor=red,urlcolor=blue]{hyperref}
\usepackage{appendix}
\usepackage{float}
\usepackage{booktabs}
\usepackage[export]{adjustbox}

\newcommand{\RNum}[1]{\uppercase\expandafter{\romannumeral #1\relax}}

\begin{document}
\title{Deformed Symmetry Structures and Quantum Many-body Scar Subspaces}
\author{Jie Ren}
\email{jieren@iphy.ac.cn}
\affiliation{Beijing National Laboratory for Condensed Matter Physics and Institute of Physics, Chinese Academy of Sciences, Beijing 100190, China}
\affiliation{University of Chinese Academy of Sciences, Beijing 100049, China}
\author{Chenguang Liang}
\affiliation{Beijing National Laboratory for Condensed Matter Physics and Institute of Physics, Chinese Academy of Sciences, Beijing 100190, China}
\affiliation{University of Chinese Academy of Sciences, Beijing 100049, China}
\author{Chen Fang}
\affiliation{Beijing National Laboratory for Condensed Matter Physics and Institute of Physics, Chinese Academy of Sciences, Beijing 100190, China}
\affiliation{Songshan Lake Materials Laboratory, Dongguan, Guangdong 523808, China}
\affiliation{Kavli Institute for Theoretical Sciences, Chinese Academy of Sciences, Beijing 100190, China}

\begin{abstract} 
A quantum many-body scar system usually contains a special non-thermal subspace (approximately) decoupled from the rest of the Hilbert space. 
In this work, we propose a general structure called deformed symmetric spaces for the decoupled subspaces hosting quantum many-body scars, which are irreducible sectors of simple Lie groups transformed by matrix-product operators (or projected entangled pair operators), of which the entanglement entropies are proved to obey sub-volume-law scaling and thus violate the eigenstate thermalization hypothesis.
A deformed symmetric space, in general, is required to have at least a U(1) sub-Lie-group symmetry to allow coherent periodic dynamics from certain low-entangled initial states.
We enumerate several possible deforming transformations based on the sub-group symmetry requirement and recover many existing models whose scar states are not connected by symmetry.
In particular, a two-dimensional scar model is proposed, which hosts a periodic dynamical trajectory on which all states are topologically ordered.
\end{abstract}

\maketitle

\section{Introduction}

The experimental discovery of slow thermalization dynamics in the Rydberg atom array~\cite{Rydberg,Rydberg-2,Rydberg-2D}, which is later formulated as a theoretical PXP model~\cite{PXP-1,PXP-2,PXP-3,PXP-4,PXP-5,PXP-6,PXP-7,PXP-su(2),TDVP,PXP-AKLT,PXP-2D-1,PXP-2D-2}, has since stimulated the study of a novel kind of weakly ergodicity breaking phenomenon, later known as the \textit{quantum many-body scar} (QMBS) (see Refs.~\cite{review,review-2} for reviews). 
The slow thermalization of the PXP model directly contrasts the strong version of the eigenstate thermalization hypothesis (ETH)~\cite{ETH-1,ETH-2,ETH-3,ETH-4,ETH-5}, which predicts that all highly-excited eigenstates in a non-integrable model locally behave like thermal ensembles, leading to the thermalization for generic initial states (with finite energy densities). 
QMBS provides explicit counter-examples where small portions of spectra, known as the \textit{scar states}, violate this hypothesis.

The origin of this non-thermal behavior in most existing QMBS systems comes from the (exact or approximate) decoupling of a special subspace $\mathcal{H}_{\mathrm{scar}}$ (called the \textit{scar space}) from the rest of the many-body Hilbert space~\cite{PXP-1,PXP-su(2),weakly-broken-Lie,unified-structure,krylov}. 
Such decoupling implies the (approximately) direct-sum form of the scar Hamiltonian:
\begin{equation}
	\hat H = \hat H_{\mathrm{scar}} \bigoplus \hat H_{\mathrm{thermal}},
	\label{decouple}
\end{equation}
and thus prevents the states in $\mathcal{H}_{\mathrm{scar}}$ from thermalization. 
Several simple models with exact decoupled scar spaces have been proposed to help understand QMBS quantitatively, including a spin-1 XY model~\cite{XY-1,XY-2}, a Rydberg-antiblockaded model~\cite{domain-wall-Tomasi,domain-wall,domain-wall-2}, a transverse field Ising ladder~\cite{Ising-ladder}, and a class of scar models in flat-band systems~\cite{flat-band,flat-band-2}. 
Remarkably, scar towers were discovered in several well-known non-integrable models, such as the Affleck-Kennedy-Lieb-Tasaki (AKLT) model~\cite{AKLT1987,AKLT-Arovas,AKLT-1,AKLT-2} and the eta-pairing Hubbard model~\cite{eta-pairing-Yang,eta-pairing-Zhang,eta-pairing-Bernevig,eta-pairing}. 

A common feature for these scar states is that they are equally spaced in energy, forming a \textit{scar tower} structure. 
In Ref.~\cite{unified-structure}, a general condition is proposed to unify many scar spaces:
\begin{equation}
	([\hat H, \hat Q^+]-\omega \hat Q^+) \mathcal{H}_{\mathrm{scar}} = 0,
	\label{eq:sga}
\end{equation}
where $\hat Q^+$ is called the \textit{raising ladder operator} of the scar tower, which generates the scar tower from an \textit{anchor state} $|\Psi_0\rangle$:
\begin{equation}
	\mathcal{H}_{\mathrm{scar}} = \mathrm{span}\{|\Psi_0\rangle, \hat Q^+|\Psi_0\rangle,(\hat Q^+)^2|\Psi_0\rangle,\cdots\}.
\end{equation}
The relation (\ref{eq:sga}) is sometimes called the \textit{restricted spectrum generating algebra} (RSGA)~\cite{SGA}. 
Based on this algebraic structure, attention shifts to the scar spaces rather than the Hamiltonians in many following works, including discovering a generalized AKLT scar model~\cite{SGA-AKLT}, a generalized $\eta$-pairing Hubbard scar model~\cite{SGA-Hubbard}, and a model whose scar tower has Onsager algebra structure~\cite{onsager}.
Moreover, it was observed that the SGA naturally occurs when $\mathcal{H}_{\mathrm{scar}}$ has a symmetry structure~\cite{qsymmetry,qsymmetry-2,qsymmetry-3,qsymmetry-4}, in which case the scar space is a symmetry sector of a Lie group $G_0$.
The scar Hamiltonian $\hat H_q$ does not have $G_0$ symmetry but is invariant in $\mathcal{H}_{\mathrm{scar}}$ under the action of $G_0$:
\begin{equation}
	\hat U_g \hat H_q|_{\mathcal H_{\mathrm{scar}}} \hat{U}^\dagger_g = \hat H_q |_{\mathcal H_{\mathrm{scar}}},\ \forall g \in G_0.
\end{equation}
This particular symmetry of subspace is called the \textit{quasisymmetry}~\cite{qsymmetry}, which explains the decoupling in some scar systems.
Furthermore, for a simple Lie group $G_0$, we can always find a U(1) subgroup of $G_0$ generated by $\hat H^z$ with equally-spaced eigenvalues.
A general scar Hamiltonian can be
\begin{equation}
	\hat H_{\mathrm{scar}} = \hat H_q + h \hat H^z, \label{eq:qsymm-ham}
\end{equation}
under which $\mathcal{H}_{\mathrm{scar}}$ is strictly decoupled, and the equally-spaced energy implies revival dynamics within $\mathcal{H}_{\mathrm{scar}}$~\cite{qsymmetry,qsymmetry-2}.

While it unifies several known exact scar models, the quasisymmetry framework still misses a number of cases. 
There are models whose scar spaces have no quasisymmetry structure (for example the AKLT model) and models having reducible symmetry sectors as their scar spaces (for example the Rydberg-blockaded model, whose scar space has U(1) quasisymmetry), in which case the degeneracies in $\hat H_q$ lack a theoretical understanding.

To address these drawbacks, in this work, we extend the previous symmetry-based theoretical framework by formulating the scar space as the \textit{deformed symmetry sector}, which is an irreducible sector of a simple Lie group $G_0$ acted by a transformation $\hat T$ preserving the non-thermal entanglement of the scar states.
A deforming transformation $\hat T$ may break the original $G_0$ symmetry but is required to preserve a subgroup symmetry $G$ (satisfying $\mathrm{U(1)} \subseteq  G \subseteq G_0$) so as to support periodic dynamics.

We adopt the notion of the matrix-product operator (MPO)~\cite{mpo-1,mpo-2} (or the projected entangled pair operator (PEPO)~\cite{tn-review} for higher dimensional systems) as the general form of such transformations.
When the dimensions of auxiliary spaces in the tensor network are finite, the deformation will preserve the sub-volume-law scaling of the entanglement entropies.
The subgroup symmetry requirement further constrains the elements of the deforming MPOs, allowing for a systematical survey for the symmetry-allowed deforming transformations.
Many previous known exact scar towers out of the scope of quasisymmetry, including the Rydberg-antiblockaded scar tower, Onsager's scar tower, AKLT scar tower, and the additional scar tower in spin-1 XY model, are successfully unified in this framework.
In addition, a new two-dimensional scar model can be constructed in this way, which supports a dynamical trajectory where all states are topologically ordered.

In this work, we focus on the scar space structures from the deformed symmetry point of view. 
We first introduce the general framework in Sec.~\ref{generalframework}, then begin systematically analyzing the deformed symmetric spaces in one-dimensional scar models in Sec.~\ref{1d-scar}.
We further discuss the newly found two-dimensional scar model with topologically ordered dynamics in Sec.~\ref{2d-scar}.
Finally, we close with the conclusion and discussion in Sec.~\ref{conclusion}.

\section{General Framework}
\label{generalframework}

\subsection{Prototype Symmetric Spaces}
\label{proto-symm}

A \textit{prototype symmetric space} $\mathcal{H}_{G_0}$ is an irreducible sector of a \textit{prototype symmetry} $G_0$, which is regarded as the underlying symmetry connecting the scar states. 
One important example is $\mathcal{H}_{\mathrm{SU(2)}}$ defined in the many-body Hilbert space of an $L$-site spin-1/2 chain, generated by
\begin{equation}
	\hat Q_{\mathrm{SU(2)}}^\pm = \sum_{j=1}^L e^{\pm ikj} \hat \sigma_j^\pm,\ 
	\hat Q_{\mathrm{SU(2)}}^z = \sum_{j=1}^L \hat \sigma_j^z.
	\label{eq:su(2)-gens}
\end{equation}
A standard basis can be constructed by sequentially applying the generator $\hat Q_{\mathrm{SU(2)}}^-$ on a fully polarized ``anchor state":
\begin{equation}
	|\Phi_{n}\rangle \equiv (\hat Q_{\mathrm{SU(2)}}^-)^{L-n} \left|\uparrow\cdots\uparrow\right\rangle.
\end{equation}
This tower structure can be extended to the multi-ladder case, where the ladder operators along with their commutation relation form a Lie algebra $\mathfrak g_0$.\footnote{A brief review of simple Lie algebras and their representations is given in Appendix~\ref{apx:lie-alg}. The ladder operators correspond to the roots of $\mathfrak g_0$, and the generalized tower corresponds to the weight system, which gives an irreducible representation of $\mathfrak g_0$.}
In general, a rank-$r$ Lie algebra $\mathfrak g_0$ is characterized by $r$ mutually commuting generators:
\begin{equation}
	\hat Q_i^z = \sum_{j=1}^N (\hat q_i^z)_j,
\end{equation}
and $r$ pairs of raising/lowering ladder operators:\footnote{Here we assume the group action is a tensor product of onsite operators: $\hat U(g) = \bigotimes_j \hat u_j(g)$. The generators of such group action are the sum of onsite operators $\hat q^\pm_i / \hat q^z_i$.}
\begin{equation}
	\hat Q_{\bm{\alpha}_i}^\pm = \sum_{j=1}^N e^{\pm i\bm{k}_{\bm \alpha_i} \cdot\bm{R}_j}(\hat{q}_{\bm \alpha_i}^\pm)_j,\label{eq:su(2)-sub-alg}
\end{equation}
where each $\bm \alpha_i$ is an $r$-dimensional vector (called the \textit{simple root}).
Each pair of ladder operators form an $\mathfrak{su}(2)$ sub-algebra:
\begin{equation}
	[\hat Q^z_i, \hat Q_{\bm \alpha_i}^\pm] = \pm \hat Q^\pm_{\bm \alpha_i},\ 
	[\hat Q_{\bm \alpha_i}^+, \hat Q_{\bm \alpha_i}^-] = 2 \hat Q_i^z.
	\label{eq:comm-rlt}
\end{equation}

An irreducible sector of $G_0$ is specified by the \textit{highest weight state} (HWS), which is chosen to be a product state: $|\Phi_{N \bm m}\rangle = \bigotimes_j |\phi_{\bm m} \rangle_j$, where $\bm m$ is an $r$-dimensional vector (called \textit{weight}) labeling the conserved quantities.
An HWS satisfies:
\begin{equation}
\begin{aligned}
	(\hat q^z_i)_j |\phi_{\bm m}\rangle_j &= m_i |\phi_{\bm m}\rangle_j, & 
	(\hat q_{\bm \alpha_i}^+)_j |\phi_{\bm m}\rangle_j &= 0, \\
	\hat Q^z_i |\Phi_{N \bm m}\rangle &= N m_i |\Phi_{N \bm m}\rangle, &
	\hat Q_{\bm \alpha_i}^+ |\Phi_{N\bm m}\rangle &= 0.
\end{aligned}
\end{equation}
All other states in the sector can be generated from HWS by applying the lowering ladder operators
\begin{equation}
\begin{aligned}
	|\Phi_{\bm M}\rangle &\equiv \hat Q_{\bm{\lambda}}^- \cdots \hat Q_{\bm{\beta}}^- \hat Q_{\bm{\alpha}}^- |\Phi_{N\bm m}\rangle, \\
	\bm M &= N\bm m - \bm \alpha - \bm \beta - \cdots -\bm \lambda.
\end{aligned}
	\label{eq:gentower}
\end{equation}
We refer to the set of states $\{|\Phi_{\bm M}\rangle\}$ as the \textit{prototype tower states}.
Together they form an orthonormal basis for $\mathcal{H}_{G_0}$.

One important property of prototype space with a (simple) Lie group symmetry is that the dimension $\dim \mathcal{H}_{G_0}$ grows asymptotically as polynomials (see Appendix~\ref{apx:lie-alg} for the proof), which is in direct contrast to the exponential growth of the many-body Hilbert space.
This moderate growth of dimension leads to the sub-volume-law bipartite entanglement of the states in $\mathcal{H}_{G_0}$. 

We note that some existing scar space can be regarded as an SU(2)-symmetric space $\mathcal{H}_{\mathrm{SU(2)}}$.
For example, the scar spaces of the spin-1 XY model in Ref.~\cite{XY-1}, eta-pairing Hubbard model in Ref.~\cite{eta-pairing}, and a spin-1/2 toy model in Refs.~\cite{PXP-su(2),eta-pairing} can all be regarded as an irreducible sector of SU(2). 
For a scar tower to be understood as an SU(2) sector, it is required that:
\begin{enumerate}
	\item The tower of states has a generating ladder operator $\hat Q^+$;
	\item The ladder operator $\hat Q^+$, together with its Hermitian conjugate $\hat Q^-$, forms an $\mathfrak{su}(2)$ algebra;
	\item The anchor state from which the tower is generated should be the highest weight state of an irreducible $\mathfrak{su}(2)$ representation.
\end{enumerate}
However, not all scar tower satisfies all three requirements.
For example, the famous AKLT model has a scar tower that starts from the AKLT ground state $|\mathrm{AKLT}\rangle$ and is generated by a ladder operator
\begin{equation}
	\hat Q^+_{\mathrm{AKLT}} = \sum_j (-1)^j (S_j^+)^2,
\end{equation}
which also correctly generates an $\mathfrak{su}(2)$ algebra.
While the anchor state $|\mathrm{AKLT}\rangle$ is not an HWS, i.e., the lowering ladder operator does not annihilate the state:
\begin{equation}
	\hat Q^-_{\mathrm{AKLT}} |\mathrm{AKLT}\rangle \ne 0.
\end{equation}
Beside, there is a Rydberg-antiblockaded scar tower generated by the ladder operator
\begin{equation}
	\hat Q_{\mathrm{Rydberg}}^+ = \sum_j (-1)^j \left(\frac{1-\hat \sigma_{j-1}^z}{2}\right) \sigma_j^+ \left(\frac{1-\hat \sigma_{j+1}^z}{2}\right).
\end{equation}
This ladder operator does satisfies $\mathfrak{su}(2)$ algebra.
Also, there are scar towers that have no ladder operators at all, for example the additional scar tower of the spin-1 XY model in Ref.~\cite{XY-1}.

\subsection{Deforming Operators}
\label{deform-mpo}

The introducing of deforming operator is motivated by the previous study of the additional tower of the spin-1 XY model~\cite{XY-2}, where the scar tower has a special structure (under periodic boundary condition):
\begin{equation}
	|\Psi_n \rangle = \hat{P}_{\mathrm{XY-II}} (\hat Q^+_{\mathrm{XY-II}})^n \bigotimes_{j=1}^{2L} \left|\downarrow\right\rangle_j,
\end{equation}
where the ``ladder-like" operator is
\begin{equation}
	\hat Q^+_{\mathrm{XY-II}} = \sum_{j=1}^L (-1)^j \sigma_{2j}^+ \sigma_{2j+1}^+,
\end{equation}
and the projective operator $\hat P_{\mathrm{XY-II}}$ is
\begin{equation}
	\hat P_{\mathrm{XY-II}} = \bigotimes_{j=1}^L \hat P_{2j-1,2j}^{(\mathrm{S=1})}.
\end{equation}
The operator $\hat P_{2j-1,2j}^{\mathrm{S=1}}$ maps two spin-1/2 degrees of freedom to a spin-1 degree of freedom:
\begin{equation}
	\hat P^{(\mathrm{S=1})} \equiv |+\rangle\left\langle\uparrow\uparrow\right| +|0\rangle (\left\langle\uparrow\downarrow\right| + \left\langle\downarrow\uparrow\right|) + |-\rangle \left\langle\downarrow\downarrow\right|.
\end{equation}
The lesson from this construction is that some seemingly complicated tower structures may have a ``hidden symmetry" (for the spin-1 XY model, the additional tower has a hidden SU(2) symmetry).

In our framework, we generalized the idea to propose a \textit{deformed symmetric space}, which is spanned by the tower state deformed by a deforming operator (denoted as $\hat T$):
\begin{equation}
	\mathcal{H}_d \equiv \mathrm{span}\{|\Psi_{\bm M}\rangle = \hat T |\Phi_{\bm M}\rangle,\ \forall \bm M\}.
\end{equation}
Although the deforming operators are simply projective in many cases, we do not impose any constraint on them so as to capture the most general cases.

One property we want the deformed tower states to preserve is the sub-volume-law entanglement entropy.
For this purpose $\hat T$ in this work is chosen to be an MPO (or a PEPO for higher dimensional cases), in which case the operator has a matrix product structure:\footnote{Note that in this work, we mainly deal with an unusual form of MPO with a ``periodic boundary condition" (the usual form of MPO is in open boundary condition, i.e., it does not take the trace). The deforming MPO is chosen in this form to preserve the periodic boundary condition when applied to homogeneous MPS.}
\begin{eqnarray}
	\hat T &=& \sum_{\bm i,\bm j} \mathrm{Tr}\left[ W^{[1]}_{i_1j_1} \cdot W^{[2]}_{i_2 j_2} \cdot \dotsc \cdot W^{[L]}_{i_L j_L} \right] \nonumber \\
	&& \times |i_1,\dots,i_L\rangle\langle j_1,\dots,j_L| \label{eq:MPO}
\end{eqnarray}
where each $W^{[k]}_{i_kj_k}$ gives a matrix of size $\chi \times \chi$, and the $\chi$-dimensional vector space is usually called the \textit{auxiliary space} of the MPO.
Each $W^{[k]}$ can then be regarded as a rank-4 tensor, whose tensor element is denoted as $W^{[k]}_{ij;\alpha\beta}$.\footnote{As a convention, we use Latin letters to represent the physical indices and Greek letters to represent the auxiliary degrees of freedom.}

Although the structure of the symmetric sector can be altered in a significant way, the tensor product structure of $\hat T$ ensures the sub-volume-law scaling of the entanglement entropy of the deformed tower.
In Appendix~\ref{apx:entropy}, we show that for a contiguous bipartition of the system (in any dimension), the entanglement entropy of an MPO/PEPO-deformed tower state scales at most as 
\begin{equation}
	S \sim O(\log V) + O(A),
\end{equation}
where $V, A$ is the volume and surface area of the sub-region.
This sub-volume-law scaling violates the ETH once those deformed states become highly excited eigenstates of a non-integrable Hamiltonian.

The MPO language also gives an MPS description of the deformed tower states. 
When acting the deforming MPO $\hat T$ on the HWS, the deformed HWS becomes
\begin{widetext}
\begin{eqnarray}
	|\Psi_{L \bm m}\rangle
	&\equiv & \hat T |\Phi_{L\bm m}\rangle \nonumber \\
	&=& \sum_{\bm i,\bm j} \mathrm{Tr}\left[ W^{[1]}_{i_1j_1} \cdot W^{[2]}_{i_2 j_2} \cdot \dotsc \cdot W^{[L]}_{i_L j_L} \right] |i_1,\dots,i_L\rangle\langle j_1,\dots,j_L| \bigotimes_{k=1}^L |\phi_{\bm m}\rangle_k \nonumber \\
	&= & \mathrm{Tr}\left[\prod_{k=1}^L\left(\sum_{j_k} W^{[k]}_{i_k j_k}\langle j_k|\phi_{\bm m}\rangle\right)\right] |i_1,\cdots,i_L\rangle \nonumber \\
	&\equiv & \mathrm{Tr}\left[A^{[1]}_{i_1}\cdot \dotsc \cdot A^{[L]}_{i_L}\right] |i_1,\cdots,i_L\rangle
\end{eqnarray}
\end{widetext}
where the tensor $A^{[k]}$ can be understood as the contraction of the MPO tensor $W$ and the local HWS $|\phi_{\bm m}\rangle$:
\begin{eqnarray}
	A^{[k]}_i \equiv \sum_j W^{[k]}_{i j} \langle j|\phi_{\bm m}\rangle.
\end{eqnarray}
Furthermore, an onsite excitation in the prototype space is transformed to a ``tensor-excitation":
\begin{eqnarray}
	B^{[k]}_{i}(\bm m') &\equiv & \sum_{j} W^{[k]}_{ij} \langle j|\phi_{\bm m'}\rangle \\
	&=& \sum_{j} W^{[k]}_{ij} \langle j|\hat q_\lambda^\dagger \cdots \hat q_\alpha^\dagger |\phi_{\bm m}\rangle, \label{eq:excited-tensor}
\end{eqnarray}
where we use an additional label $\bm m'$ to keep track of the weight:
\begin{equation}
	\bm m' = \bm m-\bm \alpha - \dotsc - \bm \lambda.
\end{equation}
A general deformed tower state can thus be expressed as a superposition of MPS with ``excited tensors" $B^{[k]}(\bm m')$ on different sites.\footnote{Note that the superposition of MPSs also appears in works on the ``tangent space methods"~\cite{tangent-1,tangent-2,tangent-3}, and is helpful in representing some quasi-excitations of the AKLT model~\cite{AKLT-1,AKLT-2,SGA-AKLT}, where the excitation tensors are generated by onsite operators acting on MPS tensors. However, in our framework, the excitation tensors are not necessarily generated by local operators, making them a more general description of scar towers.}

In general, the deforming operator $\hat T$ will break the original $G_0$-symmetry of the prototype space. However, a subgroup symmetry $G \subset G_0$ will be preserved if $\hat T$ satisfies a certain ``commutation" rule: \footnote{No that here we do note require the $\hat u$ and $\hat w$ to be the same. As long as $\hat u$ and $\hat w$ are representations of the subgroup $G$, the deformed space $\hat T \mathcal{H}_{G_0}$ will be $G$-symmetric.}
\begin{equation}
	\hat u^{\otimes L}(g)\hat{T} = \hat T \hat w^{\otimes L}(g),\ \forall g \in G, \label{eq:T-comm}
\end{equation}
where $\hat u(g),\hat w(g)$ are two (not necessarily equivalent) onsite representations of $G$.
When Eq.~(\ref{eq:T-comm}) holds, the deformed symmetric space $\mathcal{H}_d$ is invariant under the action of $\hat u^{\otimes L}(g)$:
\begin{eqnarray}
	\hat u^{\otimes L}(g) \mathcal{H}_{d} &=& \hat u^{\otimes L}(g) \hat T \mathcal{H}_{G_0} \nonumber \\
	&=& \hat T \hat w^{\otimes N}(g)\mathcal{H}_{G_0} \nonumber \\
	&=& \hat T \mathcal{H}_{G_0} \nonumber \\
	&=& \mathcal{H}_d \label{eq:u-symmetric}
\end{eqnarray}
In addition, the matrix-product structure of $\hat T$ transfers (\ref{eq:T-comm}) to an onsite symmetry condition:
\begin{equation}
	\sum_{m} u_{im}(g) W^{[k]}_{mj;\alpha\beta} = \sum_{n\gamma\tau} W^{[k]}_{in;\gamma\tau} w_{nj}(g) v_{\alpha\gamma}(g) v^{\dagger}_{\tau\beta}(g), \label{eq:explicit-symm}
\end{equation}
where $v(g)$ is a (projective) representation of $G$ in the auxiliary space.
Note that the tensor $W^{[k]}$ acts like the Clebsch-Gordan coefficients that project the tensor-product representation $w \otimes v \otimes v^\dagger$ to representation $u$, i.e.,
\begin{equation}
	u(g) = W \left[w(g) \otimes v(g)\otimes v^\dagger(g)\right] W^{-1}. \label{eq:cg}
\end{equation}
Thus, the MPO tensor $W^{[k]}$ that satisfies the symmetry requirement (\ref{eq:explicit-symm}) is proportional to the Clebsch-Gordan coefficients:
\begin{equation}
	W^{[k]}_{ij;\alpha\beta} \propto (u| w v v^\dagger)^i_{j\alpha \beta},
\end{equation}
where we use the notation $(u| w v v^\dagger)$ to denote the projector from the representation $w\otimes v \otimes v^\dagger$ to representation $u$.\footnote{Note that the Clebsch-Gordan coefficients form a matrix where the first dimension is labeled by index $i$ and the second dimension is labeled by the combined index $(j,\alpha,\beta)$, which comes from the tensor product 3 representing matrix $w(g)$, $v(g)$ and $v^\dagger(g)$.}

The symmetry condition (\ref{eq:explicit-symm}) significantly reduces the possible forms of $W^{[k]}$.
In Sec.~\ref{1d-scar}, we show explicitly how this procedure produces several known scar towers beyond the scope of the previous quasisymmetry framework.

\subsection{Parent Hamiltonians}
\label{hamiltonian}

Similar to Eq.~(\ref{eq:qsymm-ham}), a general scar Hamiltonian for a given deformed symmetric space has the form
\begin{equation}
	\hat H_{\mathrm{scar}} = \hat H_d + h\hat H^z \label{eq:gen-ham},
\end{equation}
where $\hat H_d$ is degenerate in the deformed symmetric space $\mathcal{H}_d$ and $\hat H^z$ lifts the degeneracy with equally-spaced energies.
Unlike the quasisymmetry case, where the degeneracy of $\hat H_q$ is the direct result of quasisymmetry, the degenerate states of $\hat H_d$ are related by the hidden prototype symmetry group, which is usually bigger than the quasisymmetry group.

Since we impose a subgroup symmetry condition (\ref{eq:T-comm}), Eq.~(\ref{eq:u-symmetric}) shows that the deformed space is symmetric under $\hat u^{\otimes N}(g)$. The subgroup $G$ has at least a U(1) subgroup: 
\begin{equation}
	\hat u\left(g=e^{i\theta}\right) = e^{i\theta\hat q^z},\ \forall g\in U(1),
\end{equation} 
where the elements of the U(1) subgroup is parametrized as $e^{i\theta}$, and $\hat q^z$ is the generator of the U(1) group.
The U(1) symmetry lead to the conservation of the ``charge"
\begin{equation}
	\hat H^z = \sum_j \hat q^z_j,
\end{equation}
which can then be chosen as the spectrum lifting term for the deformed tower. 
The remaining task is to find a parent Hamiltonian $\hat H_d$ that is degenerate in $\mathcal{H}_d$.

In general, a set of given states can be systematically embedded into a chaotic spectrum using a projective embedding method~\cite{embed} or a so-called ``quantum inverse method"~\cite{inverse-method,inverse-method-2}. We review those two methods here and combine them with the framework.

\subsubsection{Projective Embedding Method}
Consider an $m$-site-local cluster (denote as $[\mathrm{clst}]$) in the system, where the anchor state restricted to the cluster is also a tensor-product state:
\begin{equation}
	|\Psi_{\mathrm{clst}}\rangle \equiv \bigotimes_{j \in \mathrm{clst}} |\phi_m\rangle_j.
\end{equation}
By applying the group action, the cluster also hosts an irreducible sector of $G_0$, denoted as
\begin{equation}
	\mathcal{H}_0 \equiv \mathrm{span}\{Q_{\bm{\lambda},\mathrm{clst}}^- \cdots \hat Q_{\bm{\alpha},\mathrm{clst}}^-|\Psi_{\mathrm{clst}}\rangle\},
\end{equation}
where the ladder operators are defined on the cluster: 
\begin{equation}
	\hat Q_{\mathrm{\bm{\alpha},clst}}^\pm \equiv \sum_{j \in \mathrm{clst}} e^{\pm i\bm{k}_{\bm \alpha} \cdot \bm{R}_j} (\hat q^{\pm}_{\bm\alpha})_j.
\end{equation}
Now consider a deforming MPO/PEPO with finite-dimensional auxiliary space, which can be expressed as
\begin{equation}
	\hat T = \sum_\lambda C_\lambda \hat T_{\mathrm{clst}}^{(\lambda)}\otimes \hat T_{\overline{\mathrm{clst}}}^{(\lambda)}, \label{eq:T-decomp}
\end{equation}
where the action on cluster $\hat T_{\mathrm{clst}}^{(\lambda)}$ is subjected to the auxiliary degrees of freedom on the boundary of the cluster (labeled by $\lambda$).
We can define a cluster space as
\begin{equation}
	\mathcal{H}_{\mathrm{clst}} \equiv \mathrm{span}\{ \hat T_{\mathrm{clst}}^{(\lambda)} |\Psi\rangle,\ \forall \lambda, |\Psi\rangle \in \mathcal{H}_0 \}.
\end{equation}
We denote a projection to $\mathcal{H}_{\mathrm{clst}}$ as $\hat P_{\mathrm{clst}}$.
A general non-integrable Hamiltonian with $\mathcal{H}_{\mathrm{clst}}$ as its degenerate eigenstates has the form
\begin{equation}
	\hat H_d = \sum_{i} (1-\hat{P}^{[i]}_{\mathrm{clst}}) \hat H_{\mathrm{clst}}^{[i]} (1-\hat{P}^{[i]}_{\mathrm{clst}}), \label{eq:proj-ham}
\end{equation}
where each $\hat H_{\mathrm{clst}}^{[i]}$ is a generic Hermitian operator defined on a cluster labeled by $i$.
The Hamiltonian (\ref{eq:proj-ham}), in general, has no symmetry and is non-integrable, while all states in the deformed space $\mathcal{H}_d$ are annihilated by this Hamiltonian and thus degenerate.

In Sec.~\ref{2d-scar}, we use the projective embedding method to find a 2D scar Hamiltonian for the ``topological" scar model.

\subsubsection{Quantum Inverse Method}

Given a set of $N$ states $\{|\psi_n\rangle,\ n=1,\dotsc,N\}$ (called the \textit{target states} in this method), there is another systematic framework~\cite{inverse-method,inverse-method-2} to search for their parent Hamiltonian under which those states are degenerate eigenstates.  
The input of the quantum inverse method is a set of Hermitian operator $\{\hat h_a\}$ as a basis of Hamiltonian:
\begin{equation}
	\hat H(\{J_a\}) \equiv \sum_a J_a \hat h_a,
\end{equation}
and the method return the parameters (if any) which satisfies the requirement.
The central property for the quantum inverse method is the so-called \textit{quantum covariant matrix}:
\begin{equation}
	(C_T)_{ab} \equiv \frac{1}{2} \langle \{\hat h_a, \hat h_b\} \rangle_T - \langle \hat h_a \rangle_T \langle \hat h_b \rangle_T, \label{eq:QCM}
\end{equation}
where $\langle \cdot \rangle_T$ is the expectation on target space:
\begin{equation}
	\langle \hat O \rangle_T \equiv \frac{1}{N} \sum_{n=1}^N \langle \psi_n | \hat O | \psi_n \rangle.
\end{equation}
For a given set of parameters $\{J_a\}$, the inner product
\begin{equation}
	\sum_{ab} J_a (C_T)_{ab} J_b 
\end{equation}
measures the quantum fluctuation of the target states under the Hamiltonian $\hat H(\{J_a\})$. 
The parent Hamiltonian of the target states should have zero quantum fluctuation.
To find the parameters, we only need to find the null states of matrix $C_T$, i.e.,
\begin{equation}
	\sum_b (C_T)_{ab} J_b = 0.
\end{equation}
In many cases, there are more than one null vector of $C_T$, in which case the valid Hamiltonians form a multidimensional vector space.
A Hamiltonian as the sum of all those valid terms with random coefficients will in general gives a non-integrable scar Hamiltonian of the scar tower.

In Appendix~\ref{apx:scar-ham}, we exploit the symmetries of the target space to further decompose the covariant matrix $C_T$ into irreducible representations, and explicitly compute the parent Hamiltonians for the given scar towers.

\subsection{Perfect Revival Dynamics}
\label{dynamics}

The revival dynamics can be understood as a U(1) rotation in the prototype symmetric space.
For a specific pair of ladder operators $\{\hat Q_{\bm \alpha_j}^\pm\}$, define a coherent state labeled by a complex number $\xi$:
\begin{eqnarray}
	|\Phi_\xi\rangle 
	&\equiv & e^{\xi \hat Q_{\bm \alpha_j}^-} |\Phi_{N\bm m}\rangle \\
	&=& \bigotimes_{j=1}^N \exp\left(e^{-i \bm k_{\bm{\alpha}_j}\cdot\bm{R}_j} \xi \hat{q}^-_{\bm{\alpha}_j} \right)|\phi_{\bm m}\rangle_k \\
	& \equiv & \bigotimes_{j=1}^N |\phi_{\xi}\rangle_k
\end{eqnarray}
From the commutation relation (\ref{eq:comm-rlt}), we know that
\begin{eqnarray}
|\Phi_\xi(\theta)\rangle 
	&\equiv & e^{-i\theta\hat Q_{\bm \alpha_j}^z} |\Phi_\xi\rangle \nonumber \\
	&= & e^{-i\theta\hat Q_{\bm \alpha_j}^z} \sum_n \frac{\xi^n}{n!} |\Phi_{N\bm m - n\bm{\alpha}_j}\rangle \nonumber \\
	&\propto & \sum_n \frac{\xi^n e^{in\theta}}{n!} |\Phi_{N\bm m - n\bm \alpha_j}\rangle,
\end{eqnarray}
When the deforming MPO/PEPO preserve a U(1) subgroup symmetry satisfying:
\begin{equation}
	e^{-i \theta \hat H^z} \hat T = \hat T e^{-i \theta \hat Q^z_{\bm \alpha_j}},
\end{equation}
The dynamics of $|\Psi_{\xi}\rangle \equiv \hat T |\Phi_\xi\rangle$ under Hamiltonian (\ref{eq:gen-ham}) is
\begin{eqnarray}
	|\Psi_{\xi}(t)\rangle &=& e^{-i(\hat H_d + h \hat H^z)t} \hat T e^{\xi \hat Q_{\bm \alpha_j}^+} |\Phi_{\bm M}\rangle \nonumber \\
	&\propto& \hat T \sum_n \frac{\xi^n e^{iht}}{n!} |\Phi_{\bm M - n\bm{\alpha}_j}\rangle \nonumber \\
	&=& |\Psi_{\xi e^{iht}}\rangle,
\end{eqnarray}
and the dynamics can be understood as a rotation of $\xi$ in the complex plane.
In addition, we can also express any states on the dynamical trajectory as an MPS, with the tensor
\begin{equation}
	A^{[k]}_i(\xi) \equiv W^{[k]}_{ij} \langle j|\phi_{\xi}\rangle_k. \label{eq:dyn-state}
\end{equation}

\section{MPO Deformed Symmetric Spaces}
\label{1d-scar}

This section focus on (homogeneous) one-dimensional spin-1/2 and spin-1 chains, where the states/operators can be expressed as MPSs/MPOs.
In the following, we investigate two special scenarios where:
\begin{enumerate}
	\item The deforming MPO break the prototype SU(2) while keeping a U(1) subgroup symmetry;
	\item The prototype SU(3) symmetry is reduced to an SO(3) subgroup symmetry.
\end{enumerate}
Then, following the symmetry condition in Eq.~(\ref{eq:explicit-symm}), we enumerate some choices of $u$, $v$ and $w$, work out solutions for the deforming MPOs, and recover existing scar models in the literature.
The general scar Hamiltonians of the towers can be constructed using the method discussed above.
In Appendix~\ref{apx:scar-ham}, we construct the scar Hamiltonian for each tower and explicitly show that the solutions cover the scar Hamiltonians in the literature.

The analysis can be easily generalized to higher dimensions using the language of the projected entangled pair states/operators.

\subsection{U(1)-conserving MPO}

For future convenience, we first introduce some notations and conventions here.
Since U(1) group only has one-dimensional irreducible representation, on a certain basis, a d-dimensional representation $D(g)$ of U(1) has the diagonal form:
\begin{equation}
	D\left(g = e^{i\theta}\right) = e^{i\theta D_1} \oplus e^{i\theta D_2} \oplus \cdots \oplus e^{i\theta D_{d}}. \label{eq:U(1)-rep}
\end{equation}
Without loss of generality, we assume the $D_i$'s are in order:
\begin{equation}
	D_1 \ge D_2 \ge \cdots \ge D_d.
\end{equation}
We denote the representation $D(g)$ as
\begin{equation}
	D = D_1 \oplus D_2 \oplus \cdots \oplus D_d.
\end{equation}
Note that for a linear representation, $D_i$ is integer, while for a projective representation, $D_i$ could take any real value.

Now we consider the $S^z$-conserving translational-invariant MPO with finite auxiliary degrees of freedom, where the representation of U(1) subgroup is
\begin{equation}
	u = S \oplus (S-1) \oplus \cdots \oplus (-S). \label{eq:U(1)-rep}
\end{equation}
For simplicity, we restrict the dimension of auxiliary space to be at most 2.
Because of the projective nature of the representation $v$, we can always multiply it by an overall phase so that 
\begin{equation}
	\sum_\alpha v_\alpha = 0,
\end{equation}
where $m$ is a real number.
Without loss of generality, we choose a local basis such that $w(g)$ is diagonal:
\begin{equation}
	w = w_1 \oplus w_2 \oplus \cdots \oplus w_{2S+1}.
\end{equation}
The symmetry condition Eq.~(\ref{eq:explicit-symm}) for U(1) subgroup requires the tensor $W_{ij;\alpha\beta}$ to be zero unless the Clebsch-Gordan coefficient $(u|w v v^\dagger) \ne 0$, which means:
\begin{equation}
	u_i - w_j = v_\alpha -v_\beta,
	\label{eq:symmreq}
\end{equation}
which means
\begin{equation}
	i+w_j-S-1=2m(\alpha-\beta). \label{eq:explicit-symmreq}
\end{equation}
We enumerate several choices of $u$, $v$ and $w$ (listed in Table~\ref{tab:u(1)-solutions}), which are labeled by type 1-6.
Each type corresponds to a family of deforming MPOs, which contain existing scar towers in the literature.

\begin{table}
	\caption{
		Some choices of $u$, $v$ and $w$ for the U(1)-symmetric MPO.
		All cases contain scar towers in the existing scar models.
	}
	\centering{}
	\setlength{\tabcolsep}{2mm}
	\begin{tabular}{ccccc}
		\hline \hline 
		Type & $u$           & $v$           & $w$      	& Tower	\tabularnewline \hline 
		1    & $\frac{1}{2} \oplus -\frac{1}{2}$ & $0$ & $\frac{1}{2} \oplus -\frac{1}{2}$ & SU(2) \tabularnewline \specialrule{0em}{1pt}{1pt}
		2    & $\frac{1}{2} \oplus -\frac{1}{2}$ & $\frac{1}{2} \oplus -\frac{1}{2}$ & $\frac{1}{2}\oplus -\frac{1}{2}$  & Rydberg 	\tabularnewline \specialrule{0em}{1pt}{1pt}
		3    & $\frac{1}{2} \oplus -\frac{1}{2}$ & $\frac{1}{2} \oplus -\frac{1}{2}$ & $\frac{3}{2}\oplus -\frac{1}{2}$  & Onsager	\tabularnewline \specialrule{0em}{1pt}{1pt}
		\hline
		4    & $1\oplus 0 \oplus -1$ & $0$ & $1 \oplus -1$  & XY-I	\tabularnewline \specialrule{0em}{1pt}{1pt}
		5    & $1\oplus 0 \oplus -1$ & $\frac{1}{2} \oplus -\frac{1}{2}$ & $1 \oplus -1$  & XY-II	\tabularnewline \specialrule{0em}{1pt}{1pt}
		6    & $1\oplus 0 \oplus -1$ & $\frac{1}{2} \oplus -\frac{1}{2}$ & $2 \oplus 0$  & AKLT	\tabularnewline \specialrule{0em}{1pt}{1pt}
		\hline \hline
	\end{tabular}
	\label{tab:u(1)-solutions}
\end{table}

The prototype SU(2)-symmetry space is generated by
\begin{equation}
	\hat Q^-_{\mathrm{SU(2)}} = \begin{cases}
		\sum_j (-1)^j \hat S_j^+ & S=\frac{1}{2} \\
		\sum_j (-1)^j (\hat S_j^+)^2 & S=1
	\end{cases}
\end{equation}
acting on the HWS defined as:
\begin{equation}
	|\Phi_0\rangle = \begin{cases}
		\bigotimes_j \left|\downarrow \right\rangle_j & S= \frac{1}{2} \\
		\bigotimes_j |-\rangle_j & S = 1
	\end{cases}.
\end{equation}
For spin-1 cases, since the prototype tower states on each site can only be $|\pm\rangle$, we restrict them to these effective 2-level degrees of freedom and express $W$ as a $(3,2;2,2)$ tensor.

In the following, we explicitly work out each type of deforming MPOs, and show the existing scar tower can be recovered in this way.
Although our framework does not rely on the existence of ladder operator, in order to compare the tower we construct with those in the literature, we discuss the sufficient condition on which a ladder operator can be found for a scar tower (see Appendix~\ref{apx:ladder-operator} for details).

\subsubsection{Type-1 Deforming MPO}
For this choice of representations, the irreducible components of $u$, $v$, $w$ are:
\begin{equation}
	\begin{aligned}
		u_i &= \frac{3}{2} - i, &i& = 1,2, \\
		v_\alpha &= 1, &\alpha & = 1,\\
		w_j &= \frac{3}{2} - j, &j& = 1,2.
	\end{aligned}
\end{equation}
The symmetry requirement Eq.~(\ref{eq:explicit-symmreq}) requires that the nonzero tensor elements satisfy:
\begin{equation}
	i-j = 0,
\end{equation}
which has the solution
\begin{equation}
	\begin{aligned}
		(i,j;\alpha,\beta) &= (1,1,1,1), \\
		(i,j;\alpha,\beta) &= (2,2,1,1).
	\end{aligned}
\end{equation}
The deforming MPO tensor $W$ is the linear combination of these two independent solutions:
\begin{equation}
	W = a \left|\downarrow\rangle \langle \downarrow \right| + b \left|\uparrow\rangle \langle \uparrow \right|.
\end{equation}
When we set the parameters as
\begin{equation}
	a = b = 1,
\end{equation}
the deforming operator becomes the identity map, and thus the deformed space is identical to the prototype space.
Such SU(2)-symmetric subspace is the scar space of toy models in Ref.~\cite{eta-pairing}. 

We then use the quantum inverse method to construct the scar Hamiltonians of the tower.
We consider all translational invariant Hamiltonians with up to three-site interaction (see Appendix~\ref{apx:spin-1/2-adj-rep} for the explicit form of these operators).
The solution of the valid Hamiltonian is obtained by diagonalizing the covariant matrix.
In Appendix~\ref{apx:su(2)-tower-ham}, we carry out the calculation explicitly.
As the result, we find 23 linearly independent terms $\{\hat H_i,\ i=1,2,\cdots,23\}$, which give the general form the the Hamiltonian:
\begin{equation}
	\hat H_\mathrm{scar} = \sum_{i=1}^{23} J_i \hat H_i + h\sum_j \hat S_j^z.
\end{equation}
The scar Hamiltonian in Ref.~\cite{eta-pairing} is included in this general form.

The coherent state with periodic dynamics is simply a product state:
\begin{equation}
	|\Psi_\xi\rangle = \bigotimes_{j=1}^{L} \frac{\left|\downarrow\right\rangle_j + \xi \left|\uparrow\right\rangle_j}{\sqrt{1 + |\xi|^2}}.
\end{equation}

\subsubsection{Type-2 Deforming MPO}
For such choice of representations, $u$, $v$, $w$ are:
\begin{equation}
	\begin{aligned}
		u_i &= \frac{3}{2} - i, &i& = 1,2, \\
		v_\alpha &= \frac{3}{2} - \alpha, &\alpha & = 1,2,\\
		w_j &= \frac{3}{2} - j, &j& = 1,2.
	\end{aligned}
\end{equation}
The symmetry requirement Eq.~(\ref{eq:explicit-symmreq}) requires the nonzero tensor elements satisfy:
\begin{equation}
	i-j = \alpha - \beta,
\end{equation}
which has the solutions:
\begin{equation}
	\begin{aligned}
		(i,j;\alpha,\beta) &= (1,1;1,1), \\
		(i,j;\alpha,\beta) &= (2,2;1,1), \\
		(i,j;\alpha,\beta) &= (1,1;2,2), \\
		(i,j;\alpha,\beta) &= (2,2;2,2), \\
		(i,j;\alpha,\beta) &= (1,2;1,2), \\
		(i,j;\alpha,\beta) &= (2,1;2,1).
	\end{aligned}
\end{equation}
The general form of tensor $W$ is the linear combination of these 6 solutions, which can be expressed as:
\begin{equation}
	W = \left[\begin{array}{cc}
		a \left|\uparrow\rangle \langle \uparrow \right| + b \left|\downarrow\rangle \langle \downarrow \right| & e \hat\sigma^+ \\
		f \hat\sigma^- & c \left|\uparrow\rangle \langle \uparrow \right| + d \left|\downarrow\rangle \langle \downarrow \right|
	\end{array}\right].
	\label{eq:Rydberg-W}
\end{equation}
Consider a special choice of parameters: 
\begin{equation}
	a=b=c=0,\ d=e=f=1. \label{eq:rydberg-coeff}
\end{equation}
This deforming MPO together with SU(2)-symmetric space generates the Rydberg antiblockaded scar tower~\cite{domain-wall} (see Appendix~\ref{apx:type-2-ladder} for the proof).

In Appendix~\ref{apx:rydberg-tower-ham}, we use the quantum inverse method to construct the translational invariant scar Hamiltonian with up to three-site interaction.
There are 14 linearly independent solutions, giving the general Hamiltonian as:
\begin{equation}
	\hat H_{\mathrm{scar}} = \sum_{i=1}^{14} J_i \hat H_i + h\sum_j \hat S_j^z.
\end{equation}
The scar Hamiltonian in Refs.~\cite{domain-wall-Tomasi,domain-wall} falls into such general form (see Appendix~\ref{apx:rydberg-tower-ham} for details).

The coherent state with periodic dynamics is an MPS:
\begin{equation}
	|\Psi_\xi\rangle = \mathrm{Tr}\left[A^{[1]}_{\xi,i_1}\cdot \dotsc \cdot A^{[L]}_{\xi,i_L}\right] |i_1,\cdots,i_L\rangle
\end{equation}
where the tensor $A_\xi$ is (unnormalized):
\begin{equation}
	A^{[j]}_{\xi,\uparrow} = \left[\begin{array}{cc} 0 & 1 \\ 0 & 0\end{array}\right],\ 
	A^{[j]}_{\xi,\downarrow} = \left[\begin{array}{cc} 0 & 0 \\ (-1)^j\xi & 1\end{array}\right].
\end{equation}
The state $|\Psi_\xi\rangle$ is precisely the Rokhsar-Kivelson initial state in Refs.~\cite{domain-wall,RK-state}.

\subsubsection{Type-3 Deforming MPO}
For such choice of representations,
\begin{equation}
	\begin{aligned}
		u_i &= \frac{3}{2} - i, &i& = 1,2, \\
		v_\alpha &= \frac{3}{2} - \alpha, &\alpha & = 1,2,\\
		w_j &= \frac{7}{2} - 2j, &j& = 1,2.
	\end{aligned}
\end{equation}
The symmetry requirement Eq.~(\ref{eq:explicit-symmreq}) requires the nonzero tensor elements satisfy:
\begin{equation}
	i-2j+2 = \alpha - \beta,
\end{equation}
which has the solutions:
\begin{equation}
	\begin{aligned}
		(i,j;\alpha,\beta) &= (2,2;1,1), \\
		(i,j;\alpha,\beta) &= (2,2;2,2), \\
		(i,j;\alpha,\beta) &= (1,2;1,2), \\
		(i,j;\alpha,\beta) &= (1,1;2,1).
	\end{aligned}
\end{equation}
The symmetry allowed tensor $W$ in this case can be parameterized as:
\begin{equation}
	W = \left[\begin{array}{cc}
		a \left|\downarrow\rangle \langle \downarrow \right| & c \sigma^+  \\
		d \left|\uparrow\rangle \langle \uparrow \right| & b \left|\downarrow\rangle \langle \downarrow \right| 
	\end{array}\right].
\end{equation}
Consider the special parameters 
\begin{equation}
	a=c=d=1,\ b=0. \label{eq:onsager-coeff}
\end{equation}
The scar tower is the ``Onsager scar tower" in Ref.~\cite{onsager} (see Appendix~\ref{apx:type-3-ladder} for the proof).

In Appendix~\ref{apx:onsager-tower-ham}, we carry out the quantum inverse method to construct the general translational invariant scar Hamiltonian with up to three-site interaction for this tower.
The result is a 7-dimensional space, which gives the general scar Hamiltonian as:
\begin{equation}
	\hat H_{\mathrm{scar}} = \sum_{i=1}^7 J_i \hat H_i + h\sum_j \hat S_j^z.
\end{equation}
The scar Hamiltonian in Ref.~\cite{onsager} is explicitly shown to satisfies the general solution (see Appendix~\ref{apx:onsager-tower-ham} for details).

The coherent state is also an MPS, with the tensor (unnormalized):
\begin{equation}
	A^{[j]}_{\xi,\uparrow} = \left[\begin{array}{cc} 0 & 1 \\ (-1)^j\xi & 0\end{array}\right],\ 
	A^{[j]}_{\xi,\downarrow} = \left[\begin{array}{cc} 0 & 0 \\ 0 & 1\end{array}\right].
\end{equation}
The state $|\Psi_\xi\rangle$ is actually the initial state in Ref.~\cite{onsager}.

\subsubsection{Type-4 Deforming MPO}
For this choice of representations:
\begin{equation}
	\begin{aligned}
		u_i &= 2 - i, &i& = 1,2,3, \\
		v_\alpha &= 0, &\alpha & = 1,\\
		w_j &= 3 - 2j, &j& = 1,2.
	\end{aligned}
\end{equation}
The symmetry requirement Eq.~(\ref{eq:explicit-symmreq}) requires the nonzero tensor elements satisfy:
\begin{equation}
	i-2j+1=0,
\end{equation}
which has the solutions:
\begin{equation}
	\begin{aligned}
		(i,j;\alpha,\beta) &= (1,1;1,1), \\
		(i,j;\alpha,\beta) &= (3,2;2,2).
	\end{aligned}
\end{equation}
The symmetry allowed tensor $W$ in this case can be parameterized as:
\begin{equation}
	W = a|+\rangle\langle+| + b|-\rangle\langle-|.
\end{equation}
By choosing the parameters 
\begin{equation}
	a= b=1, \label{eq:onsager-coeff}
\end{equation}
The deforming MPO becomes identity map and thus preserves the prototype SU(2) symmetry.
This SU(2)-quasisymmetric space corresponds to the ``type-I" scar tower in the spin-1 XY model~\cite{XY-1}.

In Appendix~\ref{apx:xy-1-tower-ham}, we explicitly search the scar Hamiltonians of the scar tower.
The solutions form a 36-dimensional space.
A general scar Hamiltonian is
\begin{equation}
	\hat H = \sum_{i=1}^{36} J_i \hat H_i+ h\sum_j \hat S_j^z.
\end{equation}
The scar Hamiltonian in Ref.~\cite{XY-2} is explicitly shown to be in the vector space (see Appendix~\ref{apx:xy-1-tower-ham} for details).

The coherent state $|\Psi_\xi\rangle$ is the product state:
\begin{equation}
	|\Psi_\xi\rangle = \bigotimes_{j=1}^{N} \frac{|-\rangle_j + (-1)^j\xi|+\rangle_j}{\sqrt{1+|\xi|^2}}.
\end{equation}
When $\xi=1$, the state $|\Psi_{\xi=1}\rangle$ is the initial state in Ref.~\cite{XY-1}.

\subsubsection{Type-5 Deforming MPO}
\label{sol-5}
For this choice of representations:
\begin{equation}
	\begin{aligned}
		u_i &= 2 - i, &i& = 1,2,3, \\
		v_\alpha &= \frac{3}{2}-\alpha, &\alpha & = 1,2, \\
		w_j &= 3 - 2j, &j& = 1,2.
	\end{aligned}
\end{equation}
The symmetry requirement Eq.~(\ref{eq:explicit-symmreq}) requires the nonzero tensor elements satisfy:
\begin{equation}
	i-2j+1=\alpha-\beta,
\end{equation}
which has the solutions:
\begin{equation}
	\begin{aligned}
		(i,j;\alpha,\beta) &= (1,1;1,1), \\
		(i,j;\alpha,\beta) &= (3,2;1,1), \\
		(i,j;\alpha,\beta) &= (1,1;2,2), \\
		(i,j;\alpha,\beta) &= (3,2;2,2), \\
		(i,j;\alpha,\beta) &= (2,2;1,2), \\
		(i,j;\alpha,\beta) &= (2,1;2,1).
	\end{aligned}
\end{equation}
Eq.~(\ref{eq:symmreq}) parameterized the tensor $W$ as:
\begin{equation}
	W = \left[\begin{array}{cc} 
	a |+\rangle\langle +| + b|- \rangle\langle -| & e |0 \rangle\langle -| \\ 
	f |0 \rangle\langle +| & c|+ \rangle\langle +| + d |- \rangle\langle -| \label{eq:type-2-1}
\end{array}\right].
\end{equation}
We choose the parameters to be:
\begin{eqnarray}
	a=d=e=f=1,\ b=c=0. \label{eq:xy-2-coeff}
\end{eqnarray}
We show that the scar tower is precisely the additional scar tower in the spin-1 XY model~\cite{XY-1,XY-2} for this choice of parameters.
This additional tower has no ladder operator.
Instead of investigating the deformed tower, we show that the dynamical initial state (\ref{eq:dyn-state}) where
\begin{equation}
	|\phi_{\xi=1}\rangle_j = \frac{|-\rangle_j + (-1)^j |+\rangle_j }{\sqrt 2}
\end{equation}
is the same as that in Ref.~\cite{XY-2}.
The dynamical initial state
\begin{equation}
	|\Psi_{\xi=1}\rangle = \hat T\bigotimes_{j=1}^N |\phi_{\xi=1}\rangle_j
\end{equation}
has the MPS representation with tensor: 
\begin{equation}
\begin{aligned}
	A^{[j]}_{+} &= \frac{1}{\sqrt 2} \left[\begin{array}{cc} (-1)^j & 0 \\ 0 & 0 \end{array} \right], \\ 
	A^{[j]}_{0} &= \frac{1}{\sqrt 2} \left[\begin{array}{cc} 0 & 1 \\ (-1)^j & 0 \end{array} \right], \\
	A^{[j]}_{-} &= \frac{1}{\sqrt 2} \left[\begin{array}{cc} 0 & 0 \\ 0 & 1 \end{array} \right],
\end{aligned}
\end{equation}
which is identical to that in in Ref.~\cite{XY-2}.

In Appendix~\ref{apx:xy-2-tower-ham}, we carry out the quantum inverse method to search for all possible translational invariant scar Hamiltonian with up to three-site interaction for the tower.
As the result, we find 5 linearly independent solutions, and the general scar Hamiltonian can be expressed as
\begin{equation}
	\hat H_{\mathrm{scar}} = \sum_{i=1}^5 J_i \hat H_i + h\sum_j \hat S_j^z
\end{equation}
The original spin-1 XY Hamiltonian~\cite{XY-2} is covered by our construction (see Appendix~\ref{apx:xy-2-tower-ham} for details).

\subsubsection{Type-6 Deforming MPO}
For this choice of representations:
\begin{equation}
	\begin{aligned}
		u_i &= 2 - i, &i& = 1,2,3, \\
		v_\alpha &= \frac{3}{2}-\alpha, &\alpha & = 1,2, \\
		w_j &= 4 - 2j, &j& = 1,2.
	\end{aligned}
\end{equation}
The symmetry requirement Eq.~(\ref{eq:explicit-symmreq}) requires the nonzero tensor elements satisfy:
\begin{equation}
	i-2j+2=\alpha-\beta,
\end{equation}
which has the solutions:
\begin{equation}
	\begin{aligned}
		(i,j;\alpha,\beta) &= (2,2;1,1), \\
		(i,j;\alpha,\beta) &= (2,2;2,2), \\
		(i,j;\alpha,\beta) &= (1,2;1,2), \\
		(i,j;\alpha,\beta) &= (1,1;2,1), \\
		(i,j;\alpha,\beta) &= (3,2;2,1).
	\end{aligned}
\end{equation}
Eq.~(\ref{eq:symmreq}) parameterized the tensor $W$ as:
\begin{equation}
	W = \left[\begin{array}{cc}
		a|0\rangle\langle-| & c|+\rangle\langle-| \\
		d|+\rangle\langle+|+e|-\rangle\langle-| & b|0\rangle\langle -|
	\end{array}\right].
\end{equation}
When the parameters are chosen to be:
\begin{eqnarray}
	a = -b = -\frac{1}{\sqrt 2},\ 
	c = -d = -e = 1, \label{eq:aklt-coeff}
\end{eqnarray}
the tower becomes the scar tower in the AKLT model~\cite{AKLT-1,AKLT-2} (see Appendix~\ref{apx:type-6-ladder} for details).
We remark that there are in addition a class of generalized AKLT scar towers~\cite{SGA-AKLT} which also fall into this general form, with the parametrization:
\begin{equation}
	a = -e = c_0,\ 
	b = c_+,\ c = d = c_-. \label{eq:gen-aklt-coeff}
\end{equation}

In Appendix~\ref{apx:aklt-tower-ham}, we carry out the quantum inverse method to search for the translational invariant scar Hamiltonian with up to two-site interaction for the AKLT scar tower.
The result is an 10-dimensional vector space.
A general scar Hamiltonian can be expressed as:
\begin{equation}
	\hat H_{\mathrm{scar}} = \sum_{i=1}^{10} J_i \hat H_i + h\sum_j \hat S_j^z.
\end{equation}
In Appendix~\ref{apx:aklt-tower-ham}, we explicitly check that such general form covers the original AKLT Hamiltonian~\cite{AKLT1987}.

The coherent state $|\Psi_\xi\rangle$ is a MPS with tensor (unnormalized):
\begin{equation}
\begin{aligned}
	A^{[j]}_{+} &= \left[\begin{array}{cc} 0 & 1 \\ (-1)^{j+1}\xi & 0 \end{array} \right], \\ 
	A^{[j]}_{0} &= \frac{1}{\sqrt 2} \left[\begin{array}{cc} -1 & 0 \\ 0 & 1 \end{array} \right], \\
	A^{[j]}_{-} &= \left[\begin{array}{cc} 0 & 0 \\ -1 & 0 \end{array} \right],
\end{aligned}
\end{equation}
We note that this MPS form of coherent state for AKLT model did not appear in the literature.
In Ref.~\cite{unified-structure}, the authors proposed the same coherent state, but with a different MPS form.
The state is constructed by writing the operator
\begin{equation}
	\exp\left(\xi \hat Q^+\right)
\end{equation}
as an (open boundary) MPO~\cite{unified-structure}:
\begin{equation}
	\sum_{\bm i, \bm j}( b_l \cdot W_{i_1j_1}\cdots W_{i_Lj_L}\cdot b_r )|i_1,\dots,i_L\rangle\langle j_1,\dots,j_L|,
\end{equation}
where 
\begin{equation}
\begin{aligned}
	W &= \left[\begin{array}{cc}
		-\mathbb I & -\xi (S^+)^2 \\
		\xi (S^+)^2 & \mathbb I
	\end{array}\right], \\
	b_l &= \left[\begin{array}{c}
		1 \\
		0
	\end{array}\right], \
	b_r = [\begin{array}{cc}
		1 & 1
	\end{array} ].
\end{aligned}
\end{equation}
Since the MPO is in open boundary form, the AKLT MPS should also be convert to open boundary form to be compatible with the MPO.
A translational-invariant MPS with periodic boundary condition can be convert into an open boundary form, but usually with an enlarged auxiliary space~\cite{mps-2}.
The open boundary MPS of the translational-invariant AKLT ground state has 4-dimensional auxiliary space, so the $\mathrm{MPO}\times \mathrm{MPS}$ has a $(2\times4=8)$-dimensional auxiliary space.

In our framework, however, we choose the deforming MPOs in periodical boundary condition.
As the result, the auxiliary dimension of the coherent state we obtained is just 2.

\subsection{SO(3)-conserving MPO}

For the SO(3)-preserving MPO, we consider a nontrivial case where $u$, $w$ form $(S=1)$-representation while $v$ form a $(S=1/2)$ representation of SO(3).
For such case, the deforming MPO can be constructed by computing the Clebsch-Gordan decomposition of SO(3) group. 
We know that the direct product of the above three SO(3) representations can be reduced to:
\begin{equation}
	1\otimes\frac{1}{2}\otimes \frac{1}{2} = 0 \oplus 1 \oplus 1 \oplus 2.
\end{equation}
It means that there are two independent solutions of $W$ for choosing such $u$, $v$, and $w$.
In the following, we work out the Clebsch-Gordan decomposition explicitly and give the general form the deforming MPO tensor $W$.

\subsubsection{Deforming MPO}
Consider a composed system of one $(S=1)$-spin and two $(S=1/2)$-spin, where the tensor product representation of SU(2) group is
\begin{equation}
	D_{1}(g) \otimes D_{\frac{1}{2}}(g) \otimes D^\dagger_{\frac{1}{2}}(g),\ \forall g \in \mathrm{SU(2)}.
\end{equation}
We know 
The explicit form of the MPO tensor $W$ can be obtained by calculating the Clebsch-Gordan coefficients of this decomposition.

To decompose the representation into irreducible ones, we first investigate the root structure and find the HWS for each irreducible sector.
The roots in this case is:
\begin{equation}
	\hat Q^\pm = \hat S^\pm_1 + \hat S_2^\pm - \hat S_3^\mp,\ 
	\hat Q^z = \hat S^z_1 + \hat S_2^z - \hat S_3^z.
\end{equation}
The HWS for $S=2$ representation is
\begin{equation}
	|S=2;Q^z=+2\rangle = \left|1,\frac{1}{2},-\frac{1}{2}\right\rangle,
\end{equation}
which generate the whole sector:
\begin{widetext}
\begin{equation}
\begin{aligned}
	|S=2;Q^z=+2\rangle &= \left|1,\frac{1}{2},-\frac{1}{2}\right\rangle, \\
	|S=2;Q^z=+1\rangle &= \frac{1}{\sqrt 2}\left|0,\frac{1}{2},-\frac{1}{2}\right\rangle + \frac{1}{2}\left|1,-\frac{1}{2},-\frac{1}{2}\right\rangle - \frac{1}{2}\left|1,\frac{1}{2},\frac{1}{2}\right\rangle, \\
	|S=2;Q^z=0\rangle &= \frac{1}{\sqrt 6}\left|-1,\frac{1}{2},-\frac{1}{2}\right\rangle-\frac{1}{\sqrt 6}\left|1,\frac{1}{2},\frac{1}{2}\right\rangle + \frac{1}{\sqrt 3}\left|0,-\frac{1}{2},-\frac{1}{2}\right\rangle -\frac{1}{\sqrt 3}\left|0,\frac{1}{2},\frac{1}{2}\right\rangle, \\
	|S=2;Q^z = -1\rangle &= \frac{1}{\sqrt 2}\left|0,-\frac{1}{2},\frac{1}{2}\right\rangle + \frac{1}{2}\left|-1,\frac{1}{2},\frac{1}{2}\right\rangle - \frac{1}{2}\left|-1,-\frac{1}{2},-\frac{1}{2}\right\rangle, \\
	|S=2;Q^z=-2\rangle &= \left|-1,-\frac{1}{2},\frac{1}{2}\right\rangle.
\end{aligned}
\end{equation}
In the $Q^z = +1$ subspace, there are two linearly independent states that are orthogonal to $|S=2,Q^z=+1\rangle$. 
These two states are the HWS for two $(S=1)$-sectors. 
We choose the first HWS as
\begin{equation}
	|S=1;Q^z=+1\rangle = \frac{1}{\sqrt 2}\left|1,-\frac{1}{2},-\frac{1}{2}\right\rangle + \frac{1}{\sqrt 2}\left|1,\frac{1}{2},\frac{1}{2}\right\rangle,
\end{equation} 
which generates the irreducible sector:
\begin{equation}
\begin{aligned}
	|S=1;Q^z = +1\rangle &= \frac{1}{\sqrt 2}\left|1,-\frac{1}{2},-\frac{1}{2}\right\rangle + \frac{1}{\sqrt 2}\left|1,\frac{1}{2},\frac{1}{2}\right\rangle, \\
	|S=1;Q^z=0\rangle &= \frac{1}{\sqrt 2}\left|0,-\frac{1}{2},-\frac{1}{2}\right\rangle + \frac{1}{\sqrt 2}\left|0,\frac{1}{2},\frac{1}{2}\right\rangle, \\
	|S=1;Q^z=-1\rangle &= \frac{1}{\sqrt 2}\left|-1,-\frac{1}{2},-\frac{1}{2}\right\rangle + \frac{1}{\sqrt 2}\left|-1,\frac{1}{2},\frac{1}{2}\right\rangle.
\end{aligned}
\end{equation}
From this irreducible sector, we can directly read off the Clebsch-Gordan coefficients:
\begin{equation}
	W^a_{11;11} = W^a_{11,22} = 
	W^a_{22;11} = W^a_{22,22} =
	W^a_{33;11} = W^a_{33,22} = \frac{1}{\sqrt 2}.
\end{equation}
Besides, there is another HWS
\begin{equation}
	|S=1;Q^z=+1\rangle = \frac{1}{\sqrt 2}\left|0,\frac{1}{2},-\frac{1}{2}\right\rangle - \frac{1}{2}\left|1,-\frac{1}{2},-\frac{1}{2}\right\rangle + \frac{1}{2}\left|1,\frac{1}{2},\frac{1}{2}\right\rangle,
\end{equation}
which generated the whole irreducible sector:
\begin{equation}
\begin{aligned}
	|S=1;Q^z=+1\rangle &= \frac{1}{\sqrt 2}\left|0,\frac{1}{2},-\frac{1}{2}\right\rangle - \frac{1}{2}\left|1,-\frac{1}{2},-\frac{1}{2}\right\rangle + \frac{1}{2}\left|1,\frac{1}{2},\frac{1}{2}\right\rangle, \\
	|S=1;Q^z=0\rangle &= \frac{1}{\sqrt 2}\left|-1,\frac{1}{2},-\frac{1}{2}\right\rangle - \frac{1}{\sqrt 2}\left|1,-\frac{1}{2},\frac{1}{2}\right\rangle, \\
	|S=1;Q^z=-1\rangle &= -\frac{1}{\sqrt 2}\left|0,-\frac{1}{2},\frac{1}{2}\right\rangle - \frac{1}{2}\left|-1,\frac{1}{2},\frac{1}{2}\right\rangle + \frac{1}{2}\left|-1,-\frac{1}{2},-\frac{1}{2}\right\rangle.
\end{aligned}
\end{equation}
From this irreducible sector, we can directly read off the Clebsch-Gordon coefficients:
\begin{equation} 
	W^a_{11;11} = W^a_{33,22} = \frac{1}{2},\ 
	W^a_{11;22} = W^a_{33,11} = -\frac{1}{2},\ 
	W^a_{12;12} = W^a_{23,12} = \frac{1}{\sqrt 2},\ 
	W^a_{21;21} = W^a_{32,21} = -\frac{1}{\sqrt 2}.
\end{equation}
\end{widetext}
Two independent MPO tensor are:
\begin{eqnarray}
	W^a = \frac{1}{\sqrt 2} \left[\begin{array}{cc}
		\hat{\mathbb I} & 0 \\
		0 & \hat{\mathbb I}
	\end{array}\right],\ 
	W^b = \frac{1}{2} \left[\begin{array}{cc}
		\hat S^z & \hat S^+ \\
		\hat S^- & -\hat S^z
	\end{array}\right].
\end{eqnarray}
A general MPO satisfying SO(3) symmetry is a linear combination of $W^a$ and $W^b$:
\begin{equation}
	W(\alpha,\beta) = \left[\begin{array}{cc}
		\alpha\hat{\mathbb I}+\beta \hat S^z & \beta \hat S^+ \\
		\beta \hat S^- & \alpha\hat{\mathbb I}-\beta \hat S^z
	\end{array}\right].
\end{equation}

\subsubsection{Deformed SU(3)-symmetric space}
We consider the case where $\alpha = \beta = 1/\sqrt 3$.
The prototype symmetry is chosen to be SU(3), generated from the HWS
\begin{equation}
	|\Phi_{(N,0)}\rangle = \bigotimes_{j=1}^N |+\rangle_j,
\end{equation}
and the ladders operators
\begin{equation}
	\hat Q_{\bm \alpha_1}^- = \sum_{j=1}^N(|0\rangle\langle +|)_j,\ 
	\hat Q_{\bm \alpha_2}^- = \sum_{j=1}^N(|-\rangle\langle 0|)_j.
\end{equation}
This choice of $W(\alpha,\beta)$ together with $\mathcal{H}_{\mathrm{SU(3)}}$ gives a deformed SU(3) symmetry sector. 

Consider a coherent state (\ref{eq:dyn-state}) of lowering operator
\begin{equation}
	\hat Q_{\bm \alpha_3}^- = [\hat Q_{\bm \alpha_2}^-, \hat Q_{\bm \alpha_1}^-]
	= \sum_{j=1}^N(|-\rangle\langle +|)_j
\end{equation}
labeled by $\xi = 1$:
\begin{equation}
	|\Phi_{\xi=1}\rangle = \bigotimes_{j=1}^N \left(\frac{|-\rangle + |+\rangle}{\sqrt 2}\right)_j,
\end{equation}
the corresponding deformed state $|\Psi_{\xi=1}\rangle = \hat T |\Phi_{\xi=1}\rangle$ has the MPS representation with tensor $A=A^{\xi=1}_j$:
\begin{equation}
	A^{[+]} = \frac{1+\sigma^z}{\sqrt 6},\ 
	A^{[0]} = \frac{1}{\sqrt 3} \sigma^x,\ 
	A^{[-]} = \frac{1-\sigma^z}{\sqrt 6}.
\end{equation}
This MPS has a projected entanglement pair interpretation and is unitarily connected to the dynamical initial state for the second scar tower in spin-1 XY model~\cite{qsymmetry}.
The preserved SO(3) symmetry acting on the $|\Psi_{\xi=1}\rangle$ is enough to generate the whole deformed space, i.e.,
\begin{equation}
	\mathcal{H}_d = \mathrm{span}\{\hat U(g) |\Psi_{\xi=1}\rangle,\ \forall g \in \mathrm{SO(3)}\},
\end{equation}
which is precisely the designed scar space for a model with SO(3) as the quasisymmetry and $|\Psi_{\xi=1}\rangle$ as the anchor state~\cite{qsymmetry}.
We remark that the SO(3) quasisymmetry is incomplete to explain the degeneracy of multiple irreducible SO(3) symmetry sectors in $\mathcal{H}_d$, while the deformed symmetry connects these sectors by an underlying SU(3) symmetry.

\section{Two-dimensional Topological Scar Model}
\label{2d-scar}

\begin{figure}
	\begin{centering}
	\includegraphics[width=.95\linewidth]{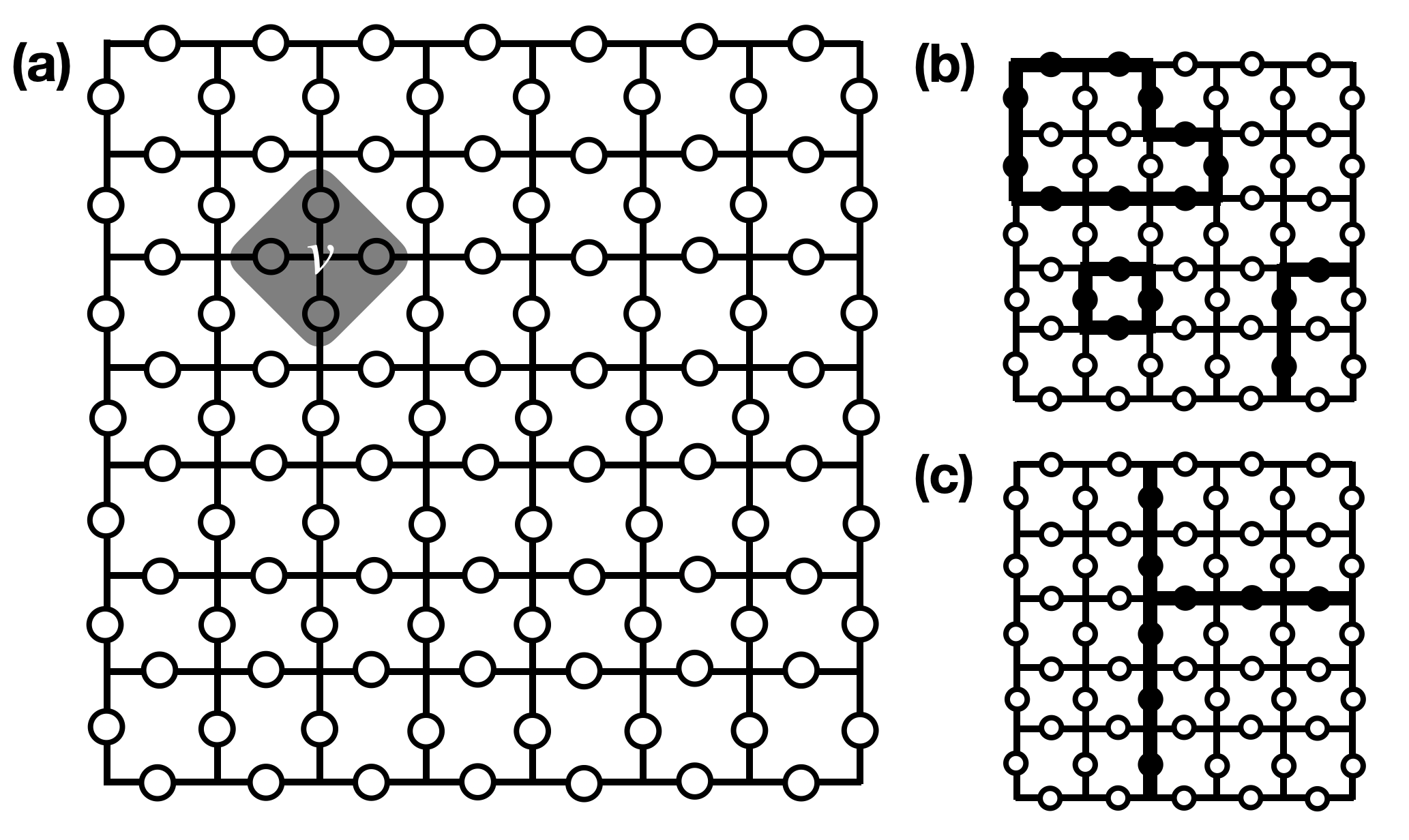}
	\par\end{centering}
	\caption{
		(a) The two-dimensional lattice system, where each spin-1/2 degree of freedom locates at the bonds, represented by a circle.
		A site $i$ directly connects to vertex $v$ is denoted as $i \in v$.
		(b)(c) Loop configuration on lattice, where each circle represents a $\left|\uparrow\right\rangle$ state while each solid dot represents a $\left|\downarrow\right\rangle$ state.
		We regard bonds with spin-down to be ``excited".
		A valid loop configuration requires that every vertex is connected to even number of excited bonds, i.e., those excited bonds form closed loops.
		(b) gives an example of valid loop configuration while (c) is not, since there is a vertex connecting to 3 excited bonds.
	}
	\label{fig:toric}
\end{figure}

Now we consider a new two-dimensional scar model constructed from this framework featuring in topologically ordered dynamical trajectory. 
We remark here that previous known ``topological scar models"~\cite{top-scar,dimer} usually have single topological states as their scar eigenstates, while in this work, we propose a model with a tower of scar states, of which specific superpositions form topologically ordered states and support periodic dynamics. 

Consider a spin-1/2 system where each spin sits on the bond of a square lattice [shown in Fig.~\ref{fig:toric} (a)], the same as the toric code model~\cite{toric-code,toric-code-2}.
We choose an SU(2) prototype symmetry generated by the usual global spin rotation.
The prototype tower states are generated from the fully polarized state
\begin{equation}
	|\Phi_0\rangle = \bigotimes_{j=1}^N \left|\uparrow \right\rangle_j
\end{equation}
acted by the ladder operator
\begin{equation}
	\hat Q^-_{\mathrm{SU(2)}} = \sum_{j=1}^N \hat S^-_j
\end{equation}
The prototype SU(2) tower states are:
\begin{equation}
	|\Phi_{n}\rangle = (\hat Q^-_{\mathrm{SU(2)}})^n |\Phi_0\rangle,
\end{equation}
which span the prototype space $\mathcal{H}_{\mathrm{SU(2)}}$.

\subsection{Projective Deforming}

Define a projection operator on the vertex $v$:
\begin{equation}
	\hat P_v = \frac{1}{2}(1+\prod_{j\in v}\hat \sigma^z_j ). \label{eq:vertex-proj}
\end{equation}
The deforming transformation $\hat T$ we use is the product of local projections on all vertices:
\begin{equation}
	\hat T = \bigotimes_v \hat P_v. \label{eq:proj-deform}
\end{equation}

Graphically, the projected states are superpositions of ``loop configurations" [as shown in Fig.~\ref{fig:toric} (b)(c)].
To show $\hat T$ is indeed a PEPO, we first express local projection $\hat P_v$ as a contracted tensor:
\begin{eqnarray}
	\hat P_v = \includegraphics[valign=c,width=.25\linewidth]{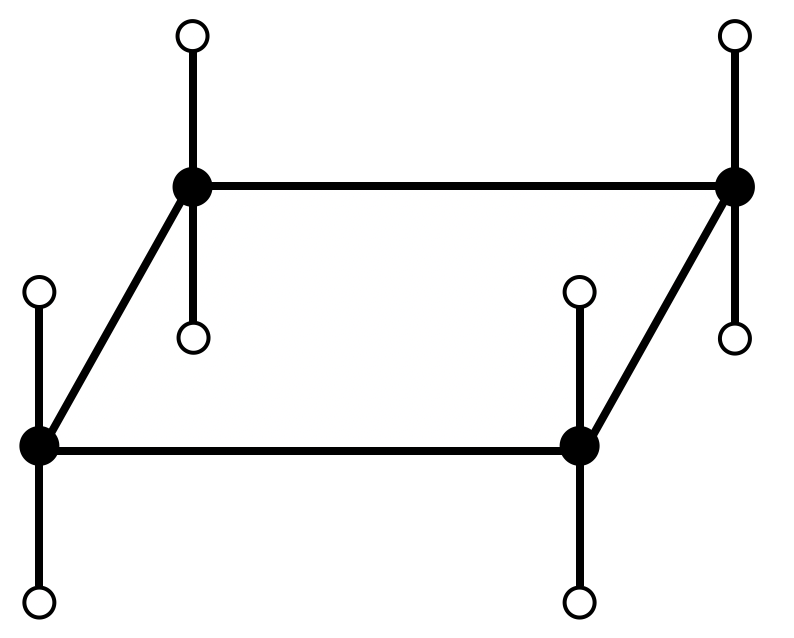},
\end{eqnarray}
where four components (corner tensors) are:
\begin{eqnarray}
	\includegraphics[valign=c,width=.5\linewidth]{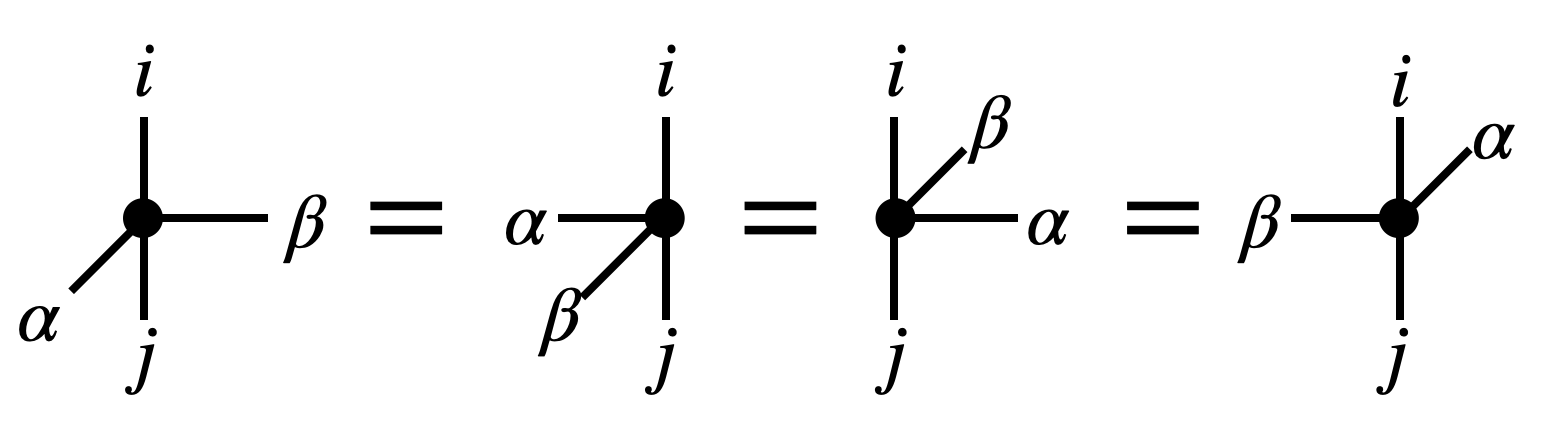} = (-1)^{i\alpha}\delta_{ij}\delta_{\alpha\beta}.
\end{eqnarray}
Operator $\hat T$ can be expressed as a tensor network of two layers of such vertex projection operators, which gives a PEPO representation.
The projection $\hat T$ breaks the original SU(2) symmetry, but a U(1) subgroup symmetry generated by
\begin{equation}
	\hat H^z = \sum_{j=1}^N \hat \sigma_j^z
\end{equation}
is preserved.

Now we consider the deformed states
\begin{equation}
	\{|\Psi_n\rangle \equiv \hat T|\Phi_n\rangle,\ \forall n\}
\end{equation}
under such projective PEPO, which can be viewed as a superposition of loop configurations.
Specifically, the state $|\Psi_n\rangle$ is the equally-weighted superposition of all loop configurations with total length $n$.\footnote{Note that not all $n$ is the total length of a valid loop configuration.
For a sufficient large lattice with $N$ spins, $n$ can only be $0,4,6,\cdots,N-6,N-4,N$
i.e., even number but excludes $2$ and $N-2$.}

\subsection{Scar Hamiltonian}

\begin{figure}
\begin{centering}
\includegraphics[width=0.8\linewidth]{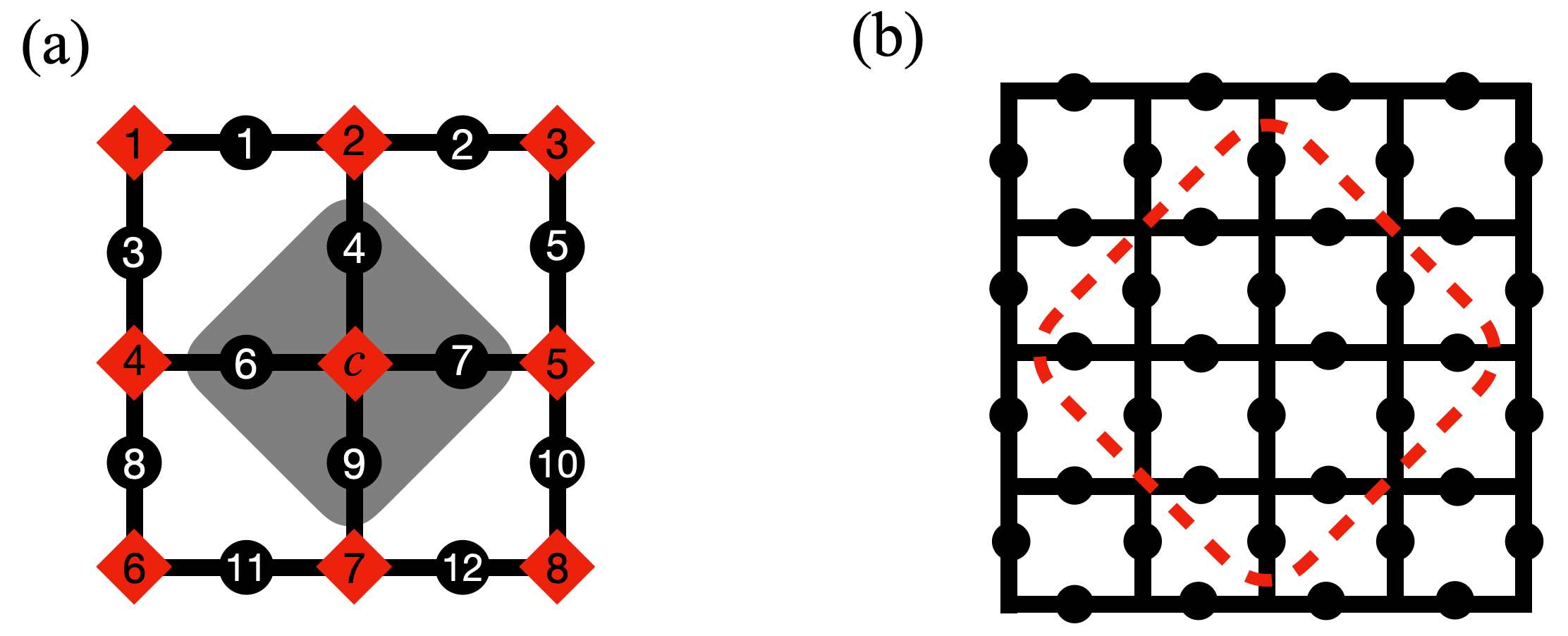}
\par\end{centering}
\caption{
	(a) Local projective cluster, where the black dots represent the spins and the red squares represent the vertices.
	In this 12-spin cluster, there are one inner vertex and 8 boundary vertices, labeled from 1 to 8.
	(b) 16-spin system in the region within the red dash line with periodic boundary condition.
}
\label{fig:projection}
\end{figure}

\begin{figure}
\begin{centering}
\includegraphics[width=0.8\linewidth]{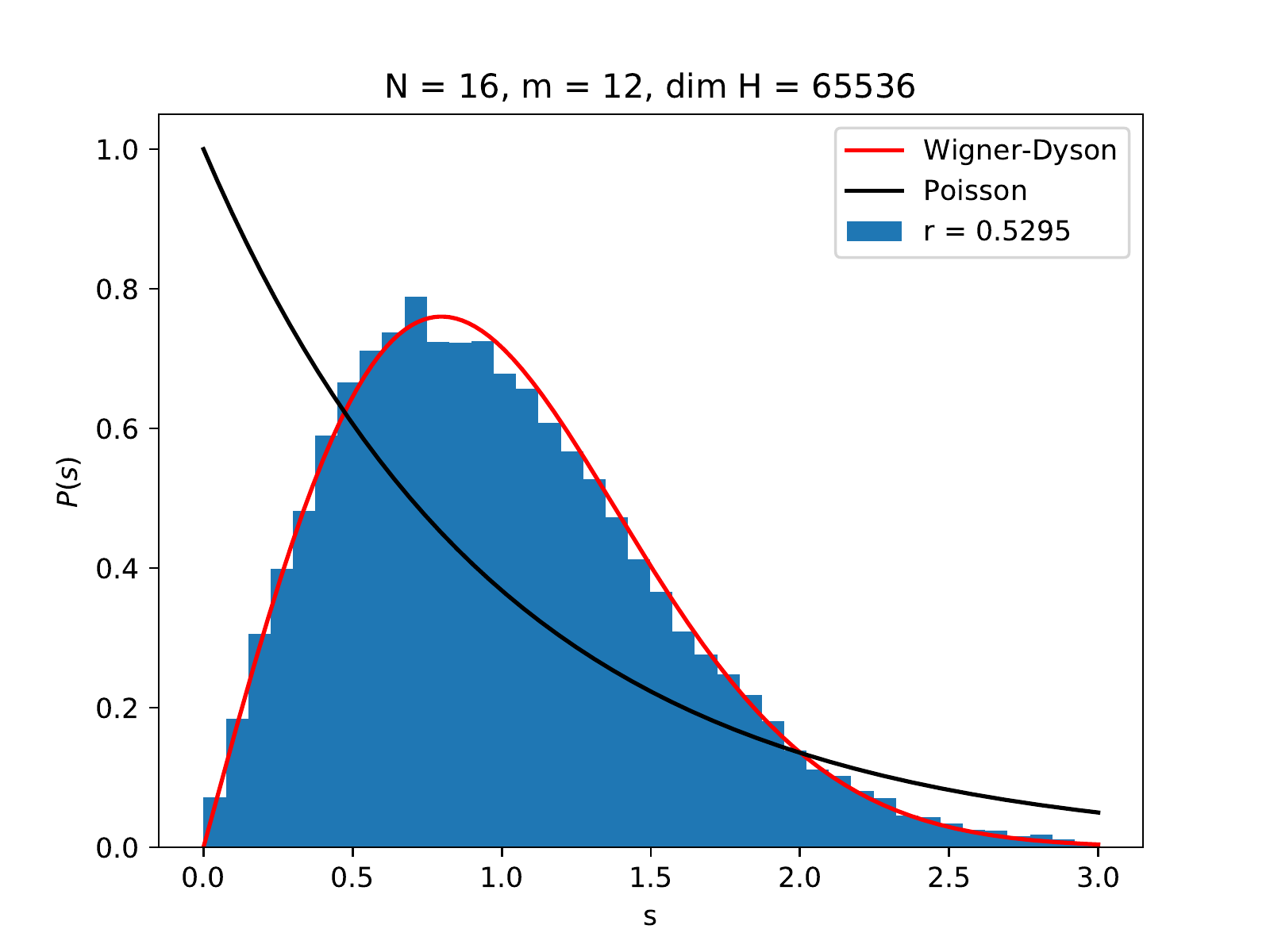}
\par\end{centering}
\caption{
	Distribution of energy level spacings in the middle half of the spectrum, where the system size $N=16$, cluster size $m=12$, and the dimension of the Hilbert space $\dim H = 65536$. 
	The r-statistics is consistent with Wigner-Dyson GOE distribution for chaotic models~\cite{level-stat}.
}
\label{fig:toric-code-level-stat}
\end{figure}

We use the cluster-projective embedding method discussed in Sec.~\ref{hamiltonian} to construct scar Hamiltonian for this set of tower states.
Consider a local cluster (denoted as $[\mathrm{clst}]$) consists of 12 spins as shown in Fig.~\ref{fig:projection} (a),
the unprojected cluster states $\{|\Phi_{\mathrm{clst},n}\rangle\}$ are 
\begin{eqnarray}
	|\Phi_{\mathrm{clst},n}\rangle &=& (\hat Q_{\mathrm{clst}}^-)^n \bigotimes_{j \in \mathrm{clst}}\left|\uparrow\right\rangle_j, \\
	\hat Q_{\mathrm{clst}}^- &=& \sum _{j\in \mathrm{clst}} \hat S_j^-.
\end{eqnarray}
To decompose the projective transformation (\ref{eq:proj-deform}) into the form (\ref{eq:T-decomp}), we first note the identity:
\begin{equation}
	\frac{1+\hat O_A \otimes \hat O_B}{2} = \sum_{p=\pm1}\left(\frac{p+\hat O_A}{2}\right)\otimes\left(\frac{p+\hat O_B}{2}\right).
	\label{eq:proj-id}
\end{equation}
We are considering the effect of the deforming operator $\hat H$ on $[\mathrm{clst}]$.
For a spin on the boundary of the cluster, the vertex projection contains the degrees of freedom within and outside the cluster:
\begin{equation}
	\hat P_v = \frac{1+\left(\prod_{j\in v_i \cap \mathrm{clst}}\sigma_j^z\right)\left(\prod_{j\in v_i \cap  \overline{\mathrm{clst}}}\sigma_j^z\right)}{2}.
\end{equation}
Using Eq. (\ref{eq:proj-id}), it can be written as:
\begin{equation}
	\hat P_v = \sum_{p_v} \left(\frac{1+\prod_{j\in v_i \cap \mathrm{clst}}\sigma_j^z}{2}\right) \left(\frac{1+\prod_{j\in v_i \cap \overline{\mathrm{clst}}}\sigma_j^z}{2}\right),
\end{equation}
where a parity $p_v=\pm 1$ is introduced to account for the correlation of the degrees of freedom inside and outside the cluster.
The deforming operator $\hat T$ in this way can be written as:
\begin{equation}
	\hat T = \sum_{\{p_v\}} \hat{T}^{\{p_v\}}_{\mathrm{clst}} \otimes \hat{T}^{\{p_v\}}_{\overline{\mathrm{clst}}},
\end{equation}
where the projection on the cluster is
\begin{equation}
	\hat T_{\mathrm{clst}}^{\{p_v\}} = \hat P_{v_c}\bigotimes_{i=1}^{8} \frac{1 + p_{v_i} \prod_{j \in v_i\cap \mathrm{clst}} \hat \sigma_j^z}{2},
\end{equation}
The cluster space is spanned by deformed cluster states under all possible boundary conditions:
\begin{equation}
	\mathcal{H}_{\mathrm{clst}} = \mathrm{span}\left\{\hat{T}_{\mathrm{clst}}^{\{p_v\}}|\Phi_{\mathrm{clst},n}\rangle,\ \forall\{p_v\}\right\}.
\end{equation}
Numerical calculation shows $\dim \mathcal{H}_{\mathrm{clst}} = 526$.
Using the cluster projection method, a random Hamiltonian $\hat{H}_0$ in Eq.~(\ref{eq:proj-ham}) can be generated.
We numerically investigate a 16 spin system [see Fig.~\ref{fig:projection} (b)] under periodic boundary condition. 
The randomly generated Hamiltonian satisfies the chaotic spectrum (Wigner-Dyson) level statistics~\cite{level-stat} (as shown in Fig.~\ref{fig:toric-code-level-stat}).

\subsection{Topologically-ordered Dynamics}

As discussed in Sec.~\ref{dynamics}, the state $|\Psi_\xi\rangle \equiv \hat T |\Phi_{\xi}\rangle$ has a close dynamical trajectory under a generic scar Hamiltonian (\ref{eq:gen-ham}) where $\hat H^z = \sum_j \hat \sigma_j^z$.
Specifically, we choose $|\Phi_{\xi=1}\rangle$ as the ($\xi = 1$) coherent state of $\hat Q^-$, then the dynamical initial state becomes: 
\begin{equation}
	|\Psi_{\xi=1}\rangle = \hat T \bigotimes_j \frac{|+\rangle_j + |-\rangle_j}{\sqrt 2} \label{eq:toric-state-init}.
\end{equation}
Note that $|\Psi_{\xi=1}\rangle$ is exactly one of the ground states of the toric code model~\cite{toric-code,toric-code-2}.
Furthermore, since the time evolution in the scar space is an onsite unitary transformation:
\begin{equation}
	\hat U(t)|_{\mathcal{H}_\mathrm{scar}} = \bigotimes_j \exp \left(-iht \hat \sigma_j^z \right),
\end{equation}
all states in the trajectory $|\Psi_{\xi=1}(t)\rangle = \hat U(t)|\Psi_{\xi=1}\rangle$ are topologically ordered as $|\Psi_{\xi=1}\rangle$.

\begin{figure}
\begin{centering}
\includegraphics[width=0.9\linewidth]{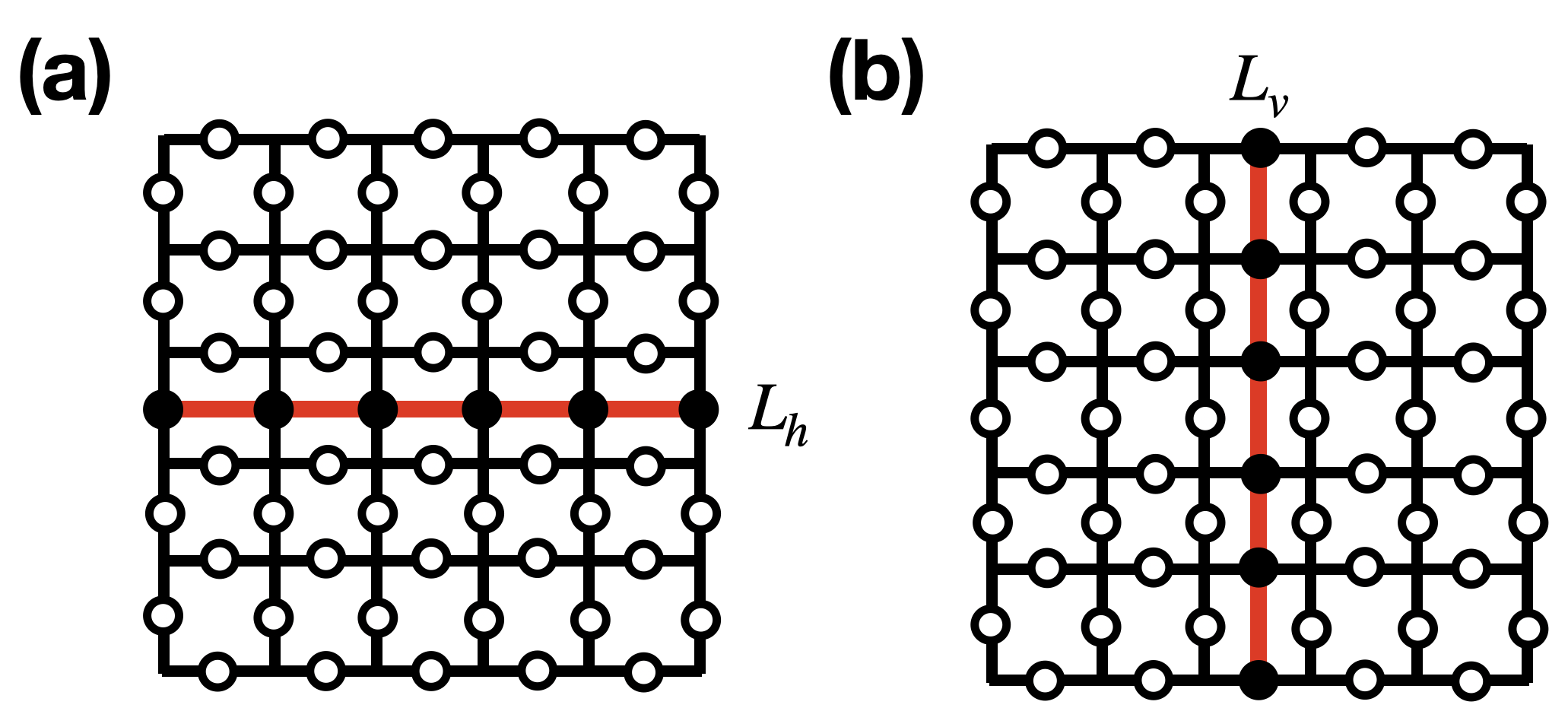}
\par\end{centering}
\caption{
	Two global loops giving topological invariants.
}
\label{fig:tc-line}
\end{figure}

The Hamiltonian constructed above supports revival topologically ordered dynamics for any ground state of the toric code model due to the topological nature of the model.
Consider two operators (cf. Fig.~\ref{fig:tc-line})
\begin{eqnarray}
	\hat Z_h = \bigotimes_{j \in L_h} \hat\sigma_j^z, \
	\hat Z_v = \bigotimes_{j \in L_v} \hat\sigma_j^z,
\end{eqnarray}
the ground states subspace of the toric code model is spanned by four orthonormal states
\begin{equation}
	\{|\Psi_{\xi=1}\rangle,\ 
	\hat Z_h |\Psi_{\xi=1}\rangle,\ 
	\hat Z_v |\Psi_{\xi=1}\rangle,\ 
	\hat Z_h \hat Z_v |\Psi_{\xi=1}\rangle \}.
\end{equation}
Note that the operator $\hat Z_h, \hat Z_v$ commute with cluster projection, so these four states are all zero-energy states of $\hat H_0$.
Remarkably, this degeneracy comes from the topological nature of these four states, which are impossible to distinguish locally.
The degeneracy also increases the dimension of the zero-energy space of $\hat H_0$.
For the 16-spin system, the deformed scar space 
\begin{equation}
	\mathcal{H}_d = \mathrm{span}\{|\Psi_n\rangle,\ n=0,4,6,8,10,12,16\}
\end{equation}
is 7-dimensional.
However, because of such topological degeneracy, the zero-energy space of $\hat H_0$ is
\begin{equation}
	\ker \hat H_0 = \mathrm{span}\{\hat Z_h^i \hat Z_h^j |\Psi_n\rangle,\ \forall i,j,n\}.
\end{equation}
The action of $\hat Z_h, \hat Z_v$ is nontrivial only when $n$ is large enough to have nontrivial loop configurations, so the dimension of $\ker \hat H_0$ is
\begin{equation}
	\dim \ker \hat H_0 = 1+4\times 5+1=22.
\end{equation}
We numerically checked that this number counting is correct.

\section{Conclusion and Discussion}
\label{conclusion}

In this work, we proposed a deformed symmetry framework to understand and construct scar models.
A general scar space can be generated from two inputs: 
(i) a prototype symmetric space and 
(ii) a deforming transformation realized by finite-dimensional MPO/PEPO.
We proved that such deformed space has a set of basis states (deformed tower state), whose bipartite entanglement entropy violates the volume-law scaling predicted by ETH.
In addition, a subgroup symmetry-preserving deformation naturally hosts periodic revival dynamics.
A parent Hamiltonian having those deformed tower states as its eigenstates can be systematically constructed using a cluster-projective embedding method. 
We investigated the possibility of the MPO tensor elements based on symmetry requirements for general MPO deforming transformations with conserved subgroup symmetries and recovered many existing scar models in this way.
In addition, a new two-dimensional scar model with a topologically ordered dynamical trajectory was constructed, where the topological nature of the ``toric-code-like" lattice was encoded into the highly excited scar eigenstates.

One open question concerns the universality of our framework, where a scar space is characterized by two symmetries: the prototype symmetry $G_0$ and the quasisymmetry $G$ (i.e., the subgroup symmetry of $G_0$ preserved by deforming transformation).
For fixed $G_0$ and $G$, the symmetry requirement gives a class of deforming MPO/PEPO transformations, which are equivalent in some ways.
A natural question is whether $G_0$ and $G$ can classify (a subset of) scar models.

Another general question concerns the possibility of generalizing the idea of deformed symmetry to those systems with only approximate decoupled scar spaces, including the original PXP model.
The quasi-periodic dynamics of the PXP model were previously studied using the time-dependent variational principle~\cite{TDVP-ref-1,TDVP-ref-2,TDVP,TDVP-2}, where the dynamical trajectory lives in a special manifold that behaves like a deformed SO(3) symmetric sector.
An interesting direction for future research is to combine the deformed symmetry framework with the variational principle to study those scar systems with approximate scar spaces theoretically and numerically.

\begin{acknowledgments}
C. F. and J. R. acknowledge support from Ministry of Science and Technology of China under Grant No. 2016YFA0302400, National Science Foundation of China under Grant No. 11674370, and Chinese Academy of Sciences under Grant No. XXH13506-202 and XDB33000000. C. L. acknowledges support from Ministry of Science and Technology of China under Grant No. 2016YFA0300600.
\end{acknowledgments}

\begin{appendix}

\section{Simple Lie Algebras and their Representations}
\label{apx:lie-alg}

A Lie algebra is called \textit{simple} (and the group it generates is called a \textit{simple Lie group}) if it is not Abelian and has no nonzero proper ideals, i.e., for a subset of a simple Lie algebra $X \subset \mathfrak{g}$,
\begin{equation}
	([A, X] \subset X,\ \forall A \in \mathfrak{g}) \ \Longrightarrow \ X = \O \mathrm{\ or\ } \mathfrak{g}.
\end{equation}
There are four types of classical algebras as the generators of:
\begin{enumerate}
	\item Special unitary groups SU($N$),
	\item Special orthogonal group SO(2N+1),
	\item Special orthogonal group SO(2N),
	\item Unitary Symplectic group USp(2N),
\end{enumerate}
and five exceptional Lie algebras $G_2$, $F_4$, $E_6$, $E_7$, and $E_8$.
In this work, we focus on the classical Lie algebras.

\subsection{Roots and Weights}

\begin{figure}
	\begin{centering}
	\includegraphics[width=.6\linewidth]{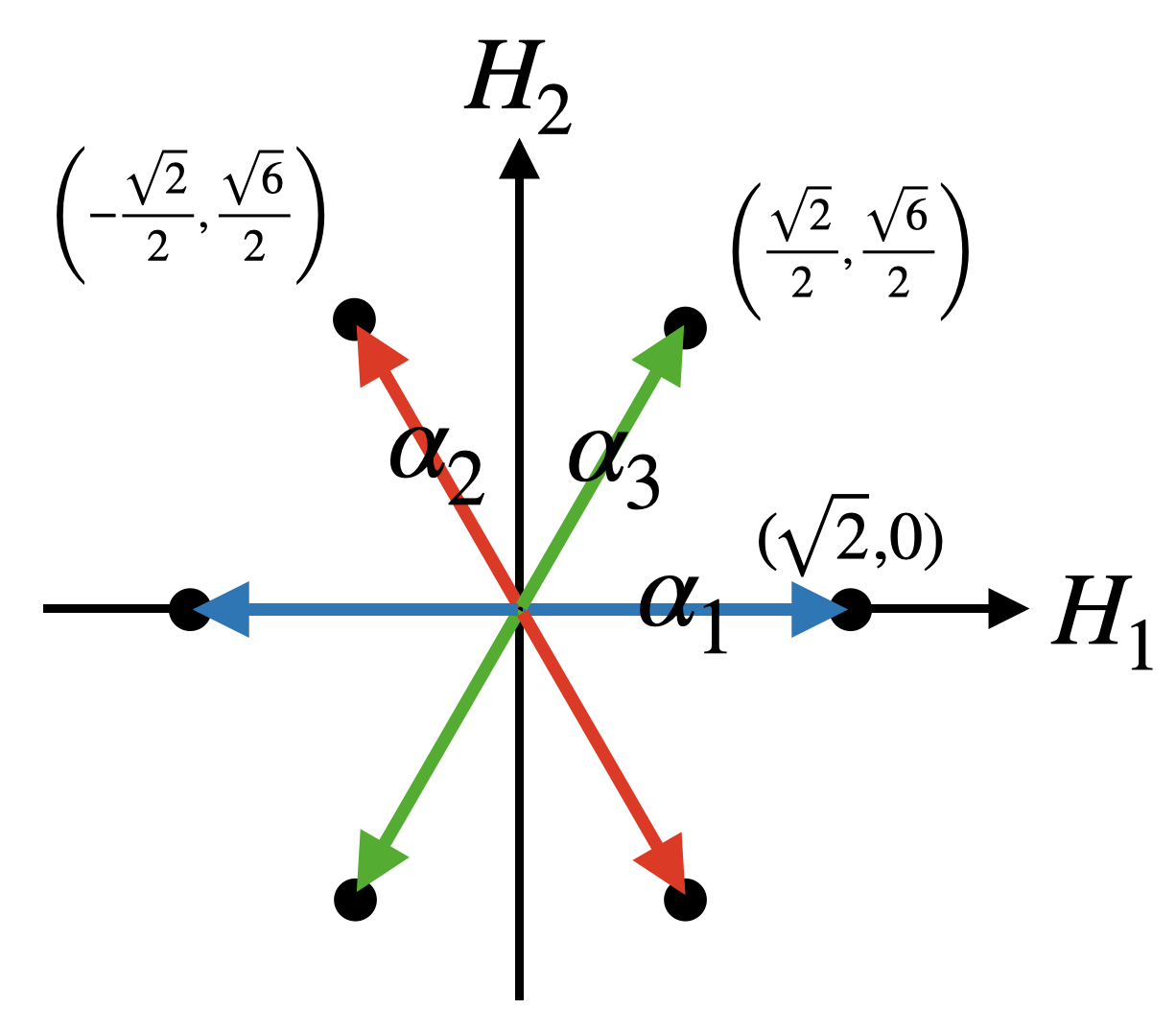}
	\par\end{centering}
	\caption{
		Roots of $\mathfrak{su}(3)$, where $\bm{\alpha}_1,\bm{\alpha}_2$ are simple roots.
	}
	\label{fig:su3-roots}
\end{figure}

For a simple Lie algebra $\mathfrak g$, a standard set of generators (which is called the \textit{Cartan-Weyl basis} of $\mathfrak g$) can be chosen so that it contains a maximal number of mutually commuting subset $\{H_i\}$ (which form the \textit{Cartan sub-algebra} of $\mathfrak g$):
\begin{equation}
	[H_i, H_j] = 0,\ \forall i,j.
\end{equation}
The maximum number of such generators $\{H_i\}$ is defined as the \textit{rank} of $\mathfrak g$.
Other generators $\{Q^\pm_{\bm \alpha}\}$ in this basis are labeled by $r$-dimensional vectors $\{\pm \bm\alpha\}$, called the \textit{roots}, which is defined by the commutation relation:
\begin{equation}
	[H_i, Q_{\bm \alpha}^\pm] = \pm \alpha_i Q^\pm_{\bm \alpha}.
\end{equation}
and the commutation relation among them is
\begin{equation}
	[Q_{\bm \alpha}, Q_{\bm \beta}] = \begin{cases}
		N_{\bm \alpha, \bm \beta} Q_{\bm \alpha + \bm \beta} & \bm \alpha + \bm \beta \mathrm{\ is\ root} \\
		\bm \alpha \cdot \bm H & \bm \alpha + \bm \beta = 0 \\
		0 & \mathrm{otherwise}
	\end{cases},
\end{equation}
where $Q_{\pm\bm{\alpha}} \equiv Q_{\bm{\alpha}}^\pm$.
For example, the $\mathfrak{su}(2)$ algebra is rank-1, with a single element Cartan sub-algebra $\{\frac{1}{\sqrt 2}\sigma^z\}$.
Other generators $\{\sigma^\pm\}$ are labeled by the root vector $\bm \alpha = (\pm1)$.
For a rank-2 $\mathfrak{su}(3)$ algebra, the Cartan sub-algebra is spanned by
\begin{equation}
\begin{aligned}
	H_1 &= \frac{1}{\sqrt 2}\left[\begin{array}{ccc}
		1 & 0 & 0 \\ 0 & -1 & 0 \\ 0 & 0 & 0
	\end{array}\right],\\ 
	H_2 &= \frac{1}{\sqrt 6}\left[\begin{array}{ccc}
		1 & 0 & 0 \\ 0 & 1 & 0 \\ 0 & 0 & -2
	\end{array}\right],
\end{aligned}
\end{equation}
and other generators are labeled by 
\begin{equation}
\begin{aligned}
	Q_{\bm{\alpha}_1}^+ &= \left[\begin{array}{ccc}
		0 & 1 & 0 \\ 0 & 0 & 0 \\ 0 & 0 & 0
	\end{array}\right], &
	Q_{\bm{\alpha}_1}^- &= \left[\begin{array}{ccc}
		0 & 0 & 0 \\ 1 & 0 & 0 \\ 0 & 0 & 0
	\end{array}\right],\\ 
	Q_{\bm{\alpha}_2}^+ &= \left[\begin{array}{ccc}
		0 & 0 & 0 \\ 0 & 0 & 1 \\ 0 & 0 & 0
	\end{array}\right], &
	Q_{\bm{\alpha}_2}^- &= \left[\begin{array}{ccc}
		0 & 0 & 0 \\ 0 & 0 & 0 \\ 0 & 1 & 0
	\end{array}\right],\\
	Q_{\bm{\alpha}_3}^+ &= \left[\begin{array}{ccc}
		0 & 0 & 1 \\ 0 & 0 & 0 \\ 0 & 0 & 0
	\end{array}\right], &
	Q_{\bm{\alpha}_3}^- &= \left[\begin{array}{ccc}
		0 & 0 & 0 \\ 0 & 0 & 0 \\ 1 & 0 & 0
	\end{array}\right],
\end{aligned}
\end{equation}
where the root vectors are (as shown in Fig.~\ref{fig:su3-roots}):
\begin{equation}
\begin{aligned}
	\bm{\alpha}_1 &= (\sqrt 2, 0),\  
	\bm{\alpha}_2 = \frac{1}{2}(-\sqrt 2, \sqrt 6),\\ 
	\bm{\alpha}_3 &= \frac{1}{2}(\sqrt 2, \sqrt 6) = \bm{\alpha}_1+\bm{\alpha}_2,
\end{aligned}
\end{equation}
The roots of $\mathfrak{su}(3)$ are not linearly independent.
We can find a set of linear independent roots called the \textit{simple roots}.
All other roots can be expressed as a linear combination of simples roots with all positive/negative coefficients. 
The $\mathfrak{su}(3)$ algebra, for example, has $\bm{\alpha}_1$ and $\bm{\alpha}_2$ as its simple roots.

\begin{figure}
	\begin{centering}
	\includegraphics[width=.6\linewidth]{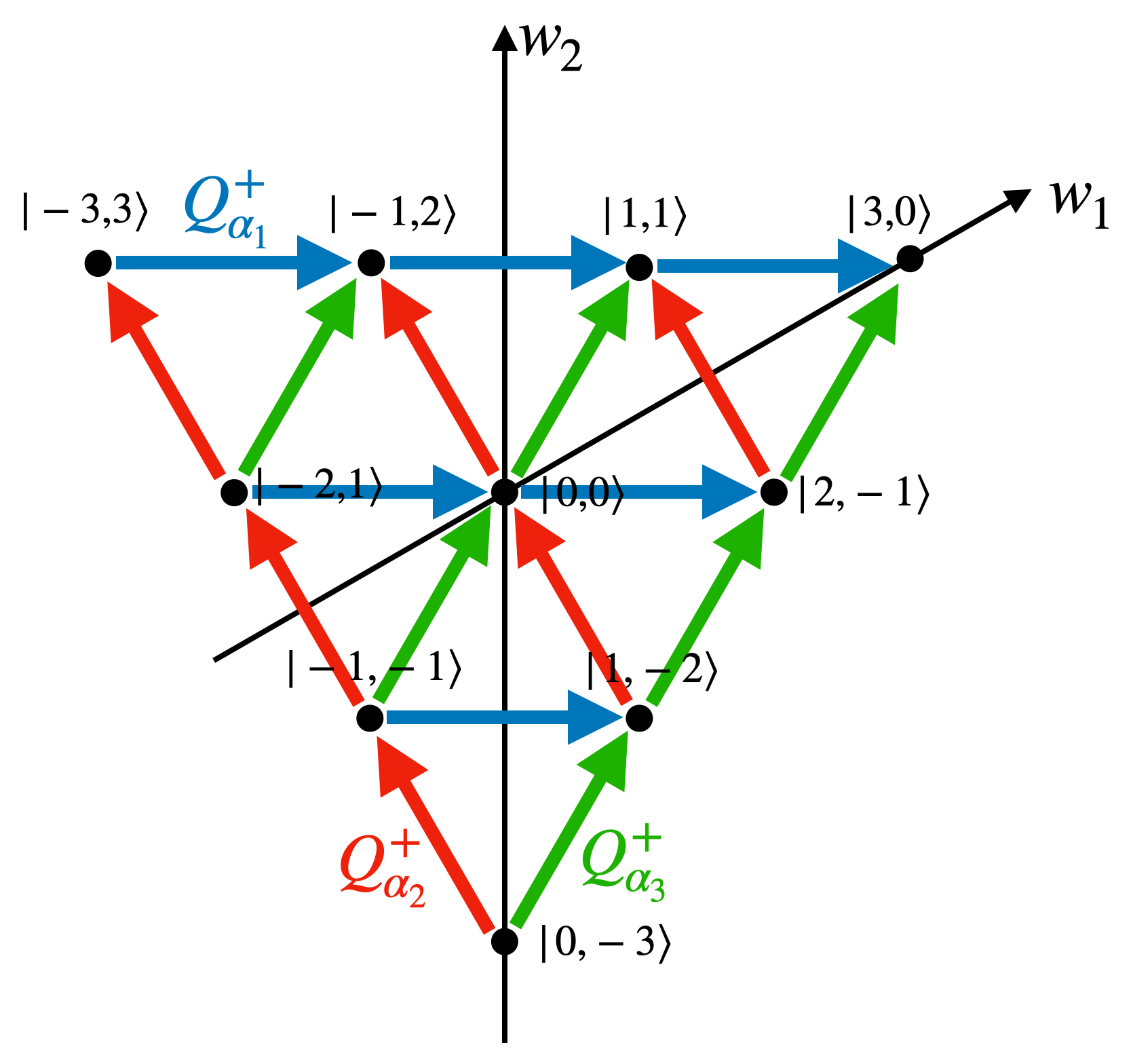}
	\par\end{centering}
	\caption{
		10 states in the irreducible SU(3)-symmetric sector generated from the HWS $|3,0\rangle$.
	}
	\label{fig:su3-weight}
\end{figure}

A \textit{state} $|\bm m\rangle$ in a representation space is labeled by an $r$-dimensional vector $\bm m$ called \textit{weight}, defined by
\begin{equation}
	H_j |\bm m\rangle = m_j |\bm m\rangle.
\end{equation}
There is a standard choice of basis vectors for weights called the \textit{fundamental weights}, defined by the orthogonal relation $\bm{\alpha}_i \cdot \bm{w}_j = \delta_{ij}$ for any simple root $\bm \alpha_i$.
For $\mathfrak{su}(3)$,
\begin{equation}
	\bm w_1 = \left(\frac{\sqrt 2}{2}, \frac{\sqrt 6}{6} \right),\ 
	\bm w_2 = \left(0,\frac{\sqrt 6}{3}\right).
\end{equation}
Under such basis, two roots of $\mathfrak{su}(3)$ becomes
\begin{equation}
	\bm \alpha_1 = (2,-1),\ \bm\alpha_2 = (-1,2).
\end{equation}
A state $|\bm M\rangle$ is called the \textit{highest weight state} (HWS) if for any simple root $\bm \alpha$, $Q^+_{\alpha}|M\rangle = 0$.
An irreducible sector of $\mathfrak g$ is specified by the highest weight $|\bm M\rangle$, with all other states generated from $|\bm M\rangle$ by multiplying generators $\{Q_{\bm \alpha_i}^-\}$:
\begin{equation}
	|\bm M'\rangle = Q^-_{\bm \alpha_n} \cdots Q^-_{\bm \alpha_2}Q^-_{\bm \alpha_1}|\bm M\rangle,
\end{equation}
where each 
\begin{equation}
	\bm M' = \bm M - \sum_{i=1}^n \bm \alpha_i
\end{equation}
is a weight in the irreducible sector. Together all $\{\bm M'\}$ form a \textit{weight system}, and the corresponding states span the representation space.
For example, an SU(3)-sector with $\bm M=(3,0)$ as the highest weight has a weight system diagrammatically shown in Fig.~\ref{fig:su3-weight}.

\subsection{Irreducible Representations}

The direct product of two HWSs $|\bm M_1\rangle, |\bm M_2\rangle$ will produce another HWS
\begin{equation}
	|\bm M_1 + \bm M_2 \rangle = |\bm M_1\rangle \otimes |\bm M_2\rangle,
\end{equation}
and $|\bm M_1 + \bm M_2 \rangle$ gives another irreducible representation.
This tensor product property of HWSs is directly related to the construction of Lie-group-symmetric many-body subspace.
In this work, the HWS is always chosen to be a product state
\begin{equation}
	|\Phi_{N\bm m}\rangle = \bigotimes_{j=1}^N |\phi_{\bm m}\rangle_j, \label{eq:prod-hws}
\end{equation}
where each $|\phi_{\bm m}\rangle_j$ is a local HWS on site $j$ labeled by weight $\bm m$.
The symmetry sector generated from $|\Phi_{N\bm M}\rangle$ corresponds to the tensor representation of Lie algebra.

In this work, for simplicity, we assume the weight $\bm m$ on each site is:
\begin{equation}
	\bm m = (1,0,\cdots,0).
\end{equation}
The tensor product of $N$ such states gives an HWS $|n\bm m\rangle$, which gives an $(N,0,\cdots,0)$ representation of $G_0$.
Mathematically, it was proved for the classical Lie algebra that the dimension of irreducible representation $(N,0,\cdots,0)$ is
\begin{widetext}
\begin{equation}
	d_{N} = \begin{dcases}
	    \displaystyle \frac{(N+n-1)(N+n-2) \cdots (N+1)}{(n-1)!} \sim O(N^{n-1}) & \mathrm{SU}(n) \\
		\displaystyle \frac{(n+2N-2)\cdot (N+n-3)(N+N-4)\cdots(N+1)}{(n-2)!} \sim O(N^{n-2})  & \mathrm{SO}(n) \\
		\displaystyle \frac{(N+2n-1)(N-2n-2)\cdots(N+1)}{(2n-1)!} \sim O(N^{2n-1}) & \mathrm{USp}(2n) \\
	\end{dcases}.
\end{equation}
\end{widetext}
We note that for a fixed $n$, $d_N$ grows (at most) polynomially with $N$ for any classical Lie algebra.
Such moderate dimension growth in contrast to the exponential growth of many-body Hilbert space is responsible for the sub-thermal properties of prototype symmetric spaces.

\section{Entanglement Entropy of Deformed Tower States}
\label{apx:entropy}

\subsection{Renyi Entropy}

For a bipartite system with Schmidt decomposition
\begin{equation}
	|\psi\rangle_{AB} = \sum_{k=1}^n \lambda_k |k\rangle_A \otimes |k\rangle_B, \label{eq:schmidt}
\end{equation}
The Renyi entropy of order $\alpha$ (for such bipartition) is defined as
\begin{equation}
	S^{(\alpha)}_A = S_B^{(\alpha)} = \frac{1}{1-\alpha} \log\left(\sum_{k=1}^n \lambda_k^{2\alpha} \right).
	\label{eq:renyi-entropy}
\end{equation}
For the case $\alpha=0,1$ where Eq.~(\ref{eq:renyi-entropy}) is not well-defined, the Renyi entropy is defined by the limit.
In particular, the zeroth-order Renyi entropy
\begin{equation}
	S^{(0)} \equiv \lim_{\alpha \rightarrow 0^+} S^{(\alpha)} = \log(n)
\end{equation}
quantifies the number of nonzero Schmidt values in the decomposition Eq.~(\ref{eq:schmidt}).
The first order Renyi entropy, also-called the \textit{von Neumann entropy}, is defined as:
\begin{equation}
	S^{(1)} \equiv \lim_{\alpha \rightarrow 1} S^{(\alpha)} = -\sum_{k=1}^n \lambda_k^2 \log(\lambda_k^2)
\end{equation}
Renyi entropies of different order satisfy the inequality:
\begin{equation}
	\alpha < \beta \Longrightarrow S^{(\alpha)} \ge S^{(\beta)}.
\end{equation}
In the following, we mainly deal with the zeroth-order Renyi entropy, as it gives the upper bound for Renyi entropy of any order.

\subsection{Entanglement Entropy of Prototype Tower States}

When decomposed into two subsystems $A,B$, the HWS becomes [as a result of (\ref{eq:prod-hws})]:
\begin{equation}
	|\Phi_{\bm M}\rangle = |\Phi_{\bm M_1}\rangle_A |\Phi_{\bm M_2}\rangle_B,
\end{equation}
where $|\Phi_{\bm M_1}\rangle_A, |\Phi_{\bm M_2}\rangle_B$ are two HWSs hosting two irreducible representations of $G_0$. 
Any prototype tower state $|\Phi_{\bm m}\rangle$ thus has the decomposition
\begin{equation}
	|\Phi_{\bm m}\rangle = \sum_{\bm m_1,\bm m_2} C^{\bm m}_{\bm m_1,\bm m_2}  |\Phi_{\bm m_1}\rangle_A |\Phi_{\bm m_2}\rangle_B,
	\label{eq:weight-decomp}
\end{equation}
where $|\Phi_{\bm m_1}\rangle_A, |\Phi_{\bm m_2}\rangle_B$ are tower states for region $A$ and $B$ respectively, and $C^{\bm m}_{\bm m_1,\bm m_2}$ is the Clebsch-Gordan coefficients.
Eq.~(\ref{eq:weight-decomp}) can be brought to the Schmidt form using the singular value decomposition, with the number of nonzero Schmidt values $N = \min\{d_{N_A},d_{N_B}\}$ (as the coefficient matrix is full rank).
The zeroth-order Renyi entropy is
\begin{equation}
	S^{(0)} = \log(\min\{d_{N_A},d_{N_B}\}) \sim O(\log V),
\end{equation}
where $V = \min\{V_A, V_B\}$ is the volume of the small region.

\subsection{Effect of Deforming Transformations}

\begin{figure}
	\begin{centering}
	\includegraphics[width=.9\linewidth]{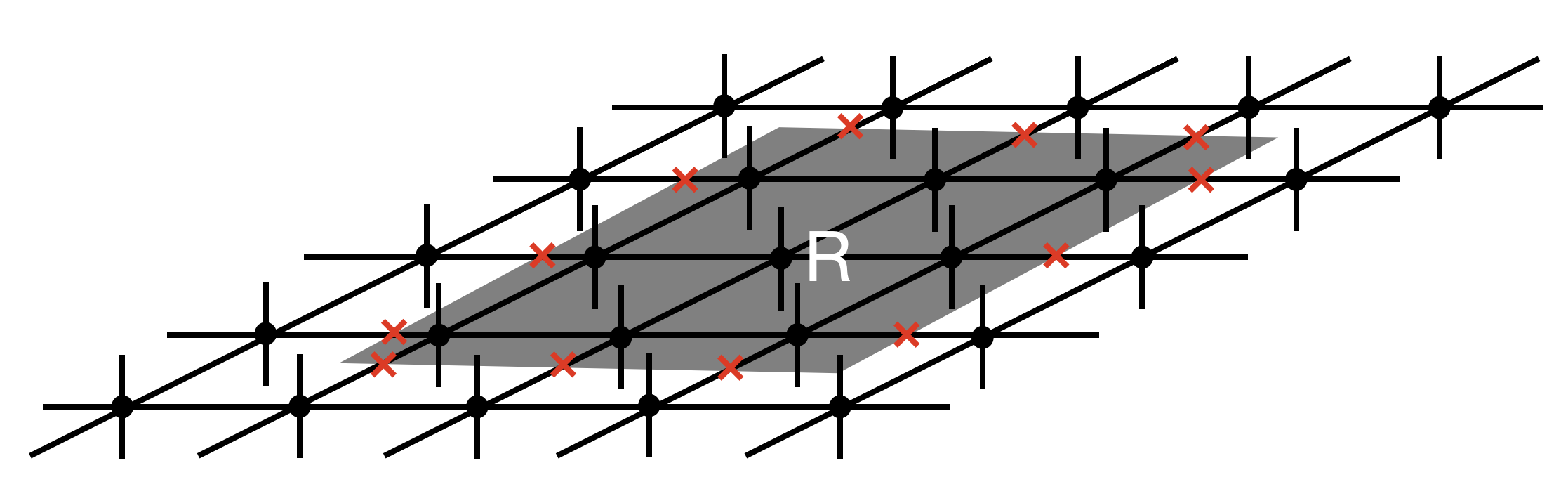}
	\par\end{centering}
	\caption{
		Decomposition of a two-dimensional PEPO, where the grey area $R$ is a contiguous region, and the red crosses mark the degrees of freedom shared by $R$ and the rest of the system $\bar R$.
		The number of the crosses is proportional to the circumference of $R$.
	}
	\label{fig:pepo}
\end{figure}

Consider an MPO (PEPO) with auxiliary dimensions less than $D$.
For a contiguous sub-region $R$ in the system (like one shown in Fig.~\ref{fig:pepo}), we can express the deforming MPO/PEPO as:
\begin{equation}
	\hat T = \sum_k C_k \hat T_R^k \otimes \hat T_{\bar R}^k
\end{equation}
The number of $k$ satisfies $N_\lambda \le D^A$, where $A$ is the ``area" of the boundary of $R$.\footnote{For a one-dimensional system, the area is a constant. For a two-dimensional system, the area is the circumference.}
Suppose the system has a Schmidt decomposition
\begin{equation}
	|\Phi\rangle = \sum_k \lambda_k |\Phi_k\rangle_R \otimes |\Phi_k\rangle_{\bar R}.
\end{equation}
When applied by $\hat T$,
\begin{equation}
	\hat T |\Phi\rangle = \sum_{k_1,k_2} C_{k_a}\lambda_{k_2} \hat T_R^{k_1}|\Phi_{k_2}\rangle_R \otimes \hat T_{\bar R}^{k_1}|\Phi_{k_2}\rangle_{\bar R}.
\end{equation}
It follows immediately that the zeroth-order Renyi entropy of the deformed state $S^{(0)}_d$ satisfies
\begin{equation}
	S^{(0)}_d \le S^{(0)} + A \log D \sim O(\log V) + O(A)
\end{equation}
where $V,A$ are the volume and surface area of the partition region.

\section{Construction of Scar Hamiltonians}
\label{apx:scar-ham}
In this appendix, we provide a systematical construction of translational-invariant scar Hamiltonians for a given tower.
The method here is based on the quantum inverse method~\cite{inverse-method,inverse-method-2}.
In principle, the parent Hamiltonians form a multi-dimensional space, which can be obtained by numerically diagonalizing the covariant matrix.
However, the result is usually ill-represented because of the freedom in choosing the basis for the vector space.
In this regard, we make use of the symmetries of the target space.
We show that if the target space is $G$-symmetric, the covariant matrix $C_T$ gives an adjoint representation of $G$, which can be block diagonalized to irreducible representations. 
The symmetry reduction of $C_T$ will simplify the diagonalizing calculation, and solutions are grouped by the symmetries.

\subsection{Adjoint Representation}
Consider the covariant matrix:
\begin{equation}
	C_T = \frac{1}{2}\langle \{\hat h_a, \hat h_b\} \rangle_T - \langle \hat h_a \rangle_T \langle \hat h_b \rangle_T.
\end{equation}
Note that in the definition of $\langle \cdot \rangle_T$, we can define the projector to the scar space as $\hat P_T$, then:
\begin{equation}
	\langle \hat O \rangle_T = \frac{1}{N}\sum_n \langle \psi_n| \hat O |\psi_n\rangle 
	= \mathrm{Tr}\left[\hat O \hat P _T\right].
\end{equation}
If the scar space process a $G$-symmetry, represented by $\hat d(g)$ in the scar space,
\begin{eqnarray}
	(C_T)_{ab} &=& \frac{1}{2}\mathrm{Tr}\left[ \{\hat d(g) \hat h_a d^\dagger(g), d(g)\hat h_b \hat d^\dagger(g)\} \hat P_T \right] -  \nonumber \\
	&&  \mathrm{Tr}\left[ \hat d(g) \hat h_a \hat d^\dagger(g) \hat P_T \right] \mathrm{Tr} \left[ \hat d(g) \hat h_b \hat d^\dagger(g)\hat P_T \right] \nonumber \\
	&=& \sum_{c,d} D^T_{ac}(g) (C_T)_{cd} D_{db}(g), \label{apx:adj-rep}
\end{eqnarray}
where $D(g)$ is the (real) adjoint representation of $G$ defined on the operator space
\begin{equation}
	\hat d(g) \hat h_a \hat d^\dagger(g) = \sum_b D_{ab}(g) \hat h_b.
\end{equation}
Eq.~(\ref{apx:adj-rep}) implies that the matrix $C_T$ can be block diagonalized in different irreducible real representations, which can further simplify the calculation.

For the $G=\mathrm{U(1)}$ case, the original one-dimensional irreducible representations are complex. 
The irreducible real representations of U(1) are the combinations of conjugate sectors, which can be denoted as $m\oplus(-m)$.

\subsection{Adjoint Representation of U(1) on Three-site Spin-1/2 Operators}
\label{apx:spin-1/2-adj-rep}
Here we regroup the translational invariant, up to three-site interacting operators into different adjoint representations of U(1). 
First, for the one-site operator, the operator space is spanned by $\{\sigma^\pm, \sigma^z\}$.
This basis also support the irreducible adjoint representation of U(1):
\begin{equation}
	e^{i\theta \hat S^z} \sigma^\pm e^{-i\theta \hat S^z} = e^{\pm i\theta} \sigma^\pm,\ 
	e^{i\theta \hat S^z} \sigma^z e^{-i\theta \hat S^z} = \sigma^z.
\end{equation}
Use the notation in (\ref{eq:U(1)-rep}), the action of U(1) spin rotation on a single site is decomposed as:
\begin{equation}
	0 \oplus (1\oplus -1).
\end{equation}
Specifically, the operator in the $0$ representation is 
\begin{equation}
	\sum_j \sigma_j^z,
\end{equation}
and there are two operator in the real representation labeled by $(1 \oplus -1)$:
\begin{equation}
\begin{aligned}
	\sum_j (\sigma_j^+ + \sigma_j^-) &= \sum_j \sigma_j^{x},\\ 
	\sum_j \frac{\sigma_j^+ - \sigma_j^-}{i} &= \sum_j \sigma_j^{y}.
\end{aligned}
\end{equation}
For the two-site operators, the decomposition of representations can be calculated by the Clebsch-Gordan series:

\begin{equation}
	(1 \oplus 0\oplus -1)^{\otimes 2}
	= (0\times 3) \oplus \left[\left(1\oplus -1\right) \times 2\right]\oplus (2\oplus -2). \label{apx:2-site-rep}
\end{equation}
Eq.~(\ref{apx:2-site-rep}) means that for two-site interacting operators, there are 3 independent operators in $0$ representation, 4 independent operators in the $1 \oplus -1$ representation, and 2 independent operator in the $2 \oplus -2$ representation.

The reduction of three-site operators can be computed in the same way:

\begin{equation}
\begin{aligned}
	(1 \oplus 0 \oplus -1)^{\otimes 3}
	&= (0\times 7) \oplus \left[\left(1\oplus -1\right) \times 6\right] \oplus \\
	&[(2\oplus -2)\times 3] \oplus \left(3\oplus -3\right).
\end{aligned}
\end{equation}
Similarly, there are 7 independent operators in $0$ representation, 12 independent operators in the $1 \oplus -1$ representation, 6 independent operator in the $2 \oplus -2$ representation, and 2 independent operator in the $3 \oplus -3$ representation.

The translational invariant operators are all of the form
\begin{equation}
	\hat h = \sum_j \hat h^{[j]}.
\end{equation}
In the $0$ representation, there are $(7 + 3\times 2+1=14)$ independent Hermitian operators, which can be further grouped into 9 reflection symmetric terms:
\begingroup
\allowdisplaybreaks
\begin{align*}
	\hat h_{0,1}^{[j]} &= \sigma_j^z, \\
	\hat h_{0,2}^{[j]} &= \sigma_j^z \sigma_{j+1}^z, \\
	\hat h_{0,3}^{[j]} &= \sigma_j^z \sigma_{j+2}^z, \\
	\hat h_{0,4}^{[j]} &= \sigma_j^z \sigma_{j+1}^z \sigma_{j+2}^z, \\
	\hat h_{0,5}^{[j]} &= \sigma_j^+ \sigma_{j+1}^- + \sigma_j^- \sigma_{j+1}^+, \\
	\hat h_{0,6}^{[j]} &= \sigma_j^+ \sigma_{j+2}^- + \sigma_j^- \sigma_{j+2}^+, \\
	\hat h_{0,7}^{[j]} &= \sigma_{j}^+ \sigma_j^z \sigma_{j+2}^- + \sigma_{j}^- \sigma_j^z \sigma_{j+2}^+, \\
	\hat h_{0,8}^{[j]} &= \sigma_j^z\sigma_{j+1}^+\sigma_{j+2}^-+\sigma_j^-\sigma_{j+1}^+\sigma_{j+2}^z+h.c., \\
	\hat h_{0,9}^{[j]} &= \frac{\sigma_j^z\sigma_{j+1}^+\sigma_{j+2}^-+\sigma_j^-\sigma_{j+1}^+\sigma_{j+2}^z}{i}+h.c.,
\end{align*}
\endgroup
and 5 reflection antisymmetric terms:
\begingroup
\allowdisplaybreaks
\begin{align*}
	\hat h_{0,10}^{[j]} &= \frac{\sigma_j^+ \sigma_{j+1}^- - \sigma_j^- \sigma_{j+1}^+}{i}, \\
	\hat h_{0,11}^{[j]} &= \frac{\sigma_j^+ \sigma_{j+2}^- - \sigma_j^- \sigma_{j+2}^+}{i}, \\
	\hat h_{0,12}^{[j]} &= \frac{\sigma_j^+ \sigma_{j+1}^z \sigma_{j+2}^- - \sigma_j^- \sigma_{j+1}^z \sigma_{j+2}^+}{i}, \\
	\hat h_{0,13}^{[j]} &= \sigma_j^z\sigma_{j+1}^+\sigma_{j+2}^- - \sigma_j^-\sigma_{j+1}^+\sigma_{j+2}^z+h.c., \\
	\hat h_{0,14}^{[j]} &= \frac{\sigma_j^z\sigma_{j+1}^+\sigma_{j+2}^- - \sigma_j^-\sigma_{j+1}^+\sigma_{j+2}^z}{i}+h.c..
\end{align*}
\endgroup
In the $(1 \oplus -1)$ representation, there are $(6\times 2+ 2\times 2\times 2 + 1\times 2 = 22)$ independent operators, which contains 14 reflection-symmetric terms:
\begingroup
\allowdisplaybreaks
\begin{align*}
	\hat h_{1,1}^{[j]} &= \sigma_j^x, \\ 
	\hat h_{1,2}^{[j]} &= \sigma_j^y, \\
	\hat h_{1,3}^{[j]} &= \sigma_j^z \sigma_{j+1}^x \sigma_{j+2}^z, \\
	\hat h_{1,4}^{[j]} &= \sigma_j^z \sigma_{j+1}^y \sigma_{j+2}^z, \\
	\hat h_{1,5}^{[j]} &= \sigma_j^z \sigma_{j+1}^x + \sigma_j^x \sigma_{j+1}^z, \\
	\hat h_{1,6}^{[j]} &= \sigma_j^z \sigma_{j+1}^y + \sigma_j^y \sigma_{j+1}^z, \\
	\hat h_{1,7}^{[j]} &= \sigma_j^z \sigma_{j+2}^x + \sigma_j^x \sigma_{j+2}^z, \\
	\hat h_{1,8}^{[j]} &= \sigma_j^z \sigma_{j+2}^y + \sigma_j^y \sigma_{j+2}^z, \\
	\hat h_{1,9}^{[j]} &= \sigma_j^z \sigma_{j+1}^z \sigma_{j+2}^x + \sigma_j^x \sigma_{j+1}^z \sigma_{j+2}^z, \\
	\hat h_{1,10}^{[j]} &= \sigma_j^z \sigma_{j+1}^z \sigma_{j+2}^y + \sigma_j^y \sigma_{j+1}^z \sigma_{j+2}^z, \\
	\hat h_{1,11}^{[j]} &= \sigma_j^- \sigma_{j+1}^+ \sigma_{j+2}^- + \sigma_j^+ \sigma_{j+1}^- \sigma_{j+2}^+, \\
	\hat h_{1,12}^{[j]} &= \frac{\sigma_j^- \sigma_{j+1}^+ \sigma_{j+2}^- - \sigma_j^+ \sigma_{j+1}^- \sigma_{j+2}^+}{i}, \\
	\hat h_{1,13}^{[j]} &= \sigma_{j}^+ \sigma_{j+1}^+ \sigma_{j+2}^- + \sigma_{j}^- \sigma_{j+1}^+ \sigma_{j+2}^+ +h.c., \\
	\hat h_{1,14}^{[j]} &= \frac{\sigma_{j}^+ \sigma_{j+1}^+ \sigma_{j+2}^- + \sigma_{j}^- \sigma_{j+1}^+ \sigma_{j+2}^+}{i} +h.c.,
\end{align*}
\endgroup
and 8 reflection antisymmetric terms:
\begingroup
\allowdisplaybreaks
\begin{align*}
	\hat h_{1,15}^{[j]} &= \sigma_j^z \sigma_{j+1}^x - \sigma_j^x \sigma_{j+1}^z, \\
	\hat h_{1,16}^{[j]} &= \sigma_j^z \sigma_{j+1}^y - \sigma_j^y \sigma_{j+1}^z, \\
	\hat h_{1,17}^{[j]} &= \sigma_j^z \sigma_{j+2}^x - \sigma_j^x \sigma_{j+2}^z, \\
	\hat h_{1,18}^{[j]} &= \sigma_j^z \sigma_{j+2}^y - \sigma_j^y \sigma_{j+2}^z, \\
	\hat h_{1,19}^{[j]} &= \sigma_j^z \sigma_{j+1}^z \sigma_{j+2}^x - \sigma_j^x \sigma_{j+1}^z \sigma_{j+2}^z, \\
	\hat h_{1,20}^{[j]} &= \sigma_j^z \sigma_{j+1}^z \sigma_{j+2}^y - \sigma_j^y \sigma_{j+1}^z \sigma_{j+2}^z, \\
	\hat h_{1,21}^{[j]} &= \sigma_{j}^+ \sigma_{j+1}^+ \sigma_{j+2}^- - \sigma_{j}^- \sigma_{j+1}^+ \sigma_{j+2}^+ +h.c., \\
	\hat h_{1,22}^{[j]} &= \frac{\sigma_{j}^+ \sigma_{j+1}^+ \sigma_{j+2}^- - \sigma_{j}^- \sigma_{j+1}^+ \sigma_{j+2}^+}{i} +h.c.,
\end{align*}
\endgroup
In the $(2 \oplus -2)$ representation, there are $(3\times 2 + 1\times 2\times 2 = 10)$ independent operators, including 8 reflection-symmetric terms:
\begingroup
\allowdisplaybreaks
\begin{align*}
	\hat h_{2,1}^{[j]} &= \sigma_j^+ \sigma_{j+1}^+ + \sigma_j^- \sigma_{j+1}^-, \\
	\hat h_{2,2}^{[j]} &= \frac{\sigma_j^+ \sigma_{j+1}^+ - \sigma_j^- \sigma_{j+1}^-}{i}, \\
	\hat h_{2,3}^{[j]} &= \sigma_j^+ \sigma_{j+2}^+ + \sigma_j^- \sigma_{j+2}^-, \\
	\hat h_{2,4}^{[j]} &= \frac{\sigma_j^+ \sigma_{j+2}^+ - \sigma_j^- \sigma_{j+2}^-}{i}, \\
	\hat h_{2,5}^{[j]} &= \sigma_j^+ \sigma_{j+1}^z\sigma_{j+2}^+ + \sigma_j^- \sigma_{j+1}^z \sigma_{j+2}^-, \\
	\hat h_{2,6}^{[j]} &= \frac{\sigma_j^+ \sigma_{j+1}^z\sigma_{j+2}^+ - \sigma_j^- \sigma_{j+1}^z \sigma_{j+2}^-}{i}, \\
	\hat h_{2,7}^{[j]} &= \sigma_j^z\sigma_{j+1}^+\sigma_{j+2}^+ + \sigma_j^+\sigma_{j+1}^+\sigma_{j+2}^z + h.c., \\
	\hat h_{2,8}^{[j]} &=\frac{\sigma_j^z\sigma_{j+1}^+\sigma_{j+2}^+ + \sigma_j^+\sigma_{j+1}^+\sigma_{j+2}^z}{i} + h.c.,
\end{align*}
\endgroup
and 2 reflection antisymmetric terms:
\begingroup
\allowdisplaybreaks
\begin{align*}
	\hat h_{2,9}^{[j]} &= \sigma_j^z\sigma_{j+1}^+\sigma_{j+2}^+ - \sigma_j^+\sigma_{j+1}^+\sigma_{j+2}^z + h.c., \\
	\hat h_{2,10}^{[j]} &=\frac{\sigma_j^z\sigma_{j+1}^+\sigma_{j+2}^+ - \sigma_j^+\sigma_{j+1}^+\sigma_{j+2}^z}{i} + h.c.,
\end{align*}
\endgroup
In the $(3 \oplus -3)$ representation, there are $(1\times 2 = 2)$ independent Hermitian operators:
\begingroup
\allowdisplaybreaks
\begin{align*}
	\hat h_{3,1}^{[j]} &= \sigma_j^+ \sigma_{j+1}^+ \sigma_{j+2}^+ + \sigma_j^- \sigma_{j+1}^- \sigma_{j+2}^-,  \\
	\hat h_{3,2}^{[j]} &= \frac{\sigma_j^+ \sigma_{j+1}^+ \sigma_{j+2}^+ - \sigma_j^- \sigma_{j+1}^- \sigma_{j+2}^-}{i}.
\end{align*}
\endgroup

For the target space with total $\hat S^z$ conserved, the covariant matrix $C_T$ is of block-diagonal form:
\begin{equation}
	C_T = C_T^{0} \oplus C_T^{1} \oplus C_T^{2} \oplus C_T^{3}.
\end{equation}
If in addition the target space is reflection-symmetric, each $C_T^m$ block can be further decomposed into different parity sector:
\begin{equation}
	C_T^m = C_T^{m+} \oplus C_T^{m-}.
\end{equation}
Searching for all independent null vector of $C_T$ is equivalent to searching the null vectors in each sector and combine the results.

\subsection{Hamiltonians for SU(2) Scar Tower}
\label{apx:su(2)-tower-ham}
We then carry out the calculation for the SU(2) tower.
The covariant matrix is computed on the $L=10$ system with periodic boundary condition.
Because of the total $\hat S^z$ and reflection symmetry, we just need to search the null vector in each $(m,p)$-sector ($m$ is the $S^z$ number and $p=\pm 1$ is the parity).

As the result, in the $(m=0,p=+1)$ sector, we find 4 independent solutions:
\begingroup
\allowdisplaybreaks
\begin{align*}
	\hat H_1 &= \hat h_{0,2} + 2\hat h_{0,5}, \\
	\hat H_2 &= \hat h_{0,3} + 2\hat h_{0,6}, \\
	\hat H_3 &= \hat h_{0,8} - 2\hat h_{0,7}, \\
	\hat H_4 &= \hat h_{0,1} - \hat h_{0, 4} - \hat h_{0,8}.
\end{align*}
\endgroup
We find that all term in the $(m=0,p=-1)$ sector is the valid solution, i.e.,
\begingroup
\allowdisplaybreaks
\begin{align*}
	\hat H_5 &= \hat h_{0,10}, \\
	\hat H_6 &= \hat h_{0,11}, \\
	\hat H_7 &= \hat h_{0,12}, \\
	\hat H_8 &= \hat h_{0,13}, \\
	\hat H_9 &= \hat h_{0,14}.
\end{align*}
\endgroup
In the $(m=1,p=+1)$ sector, there are 4 null vectors:
\begingroup
\allowdisplaybreaks
\begin{align*}
	\hat H_{10} &= \hat h_{1,1} - \hat h_{1,3} -2\hat h_{1,13}, \\
	\hat H_{11} &= \hat h_{1,2} - \hat h_{1,4} - 2\hat h_{1,14}, \\
	\hat H_{12} &= \hat h_{1,9} - 2\hat h_{1, 3} - 2 \hat h_{1,13} +4 \hat h_{1,11}, \\
	\hat H_{13} &= \hat h_{1,10} - 2\hat h_{1,4}-2\hat h_{1,14}+4\hat h_{1,12}.
\end{align*}
\endgroup
We also find that all term in the antisymmetric sectors $(m=1,p=-1)$ and $(m=2, p=-1)$ is the valid Hamiltonian, i.e.,
\begin{align*}
	\hat H_{14} &= \hat h_{1,15}, \\
	\hat H_{15} &= \hat h_{1,16}, \\
	\hat H_{16} &= \hat h_{1,17}, \\
	\hat H_{17} &= \hat h_{1,18}, \\
	\hat H_{18} &= \hat h_{1,19}, \\
	\hat H_{19} &= \hat h_{1,20}, \\
	\hat H_{20} &= \hat h_{1,21}, \\
	\hat H_{21} &= \hat h_{1,22}, \\
	\hat H_{22} &= \hat h_{2,9}, \\
	\hat H_{23} &= \hat h_{2,10}.
\end{align*}
There is no null vector in the $(m=2,p=+1)$ and $(m=3)$ sectors.
In together we find 23 linear independent Hamiltonian for the SU(2)-symmetric scar space.
The general form of the scar Hamiltonian can be
\begin{equation}
	\hat H(\{J_i\}) = \sum_{i=1}^{23} J_i \hat H_i +\hat H^z.
\end{equation}
The scar Hamiltonian in Ref.~\cite{eta-pairing} for the SU(2) tower is:
\begin{equation} 
\begin{aligned}
	\hat H_{\mathrm{SU(2)}} &= J_1\sum_j \vec \sigma_j \cdot \vec \sigma_j + J_2 \sum_j \vec \sigma_j \cdot \vec \sigma_{j+1} + \\ 
	&D \sum_j (\sigma_j^x \sigma_{j+1}^y-\sigma_j^y\sigma_{j+1}^x).
\end{aligned}
\end{equation}
The first and the second terms are nothing but $\hat H_1$ and $\hat H_2$ in our general solution, and the third term is $\hat H_5$.
This scar Hamiltonian thus falls into the general solution we have just constructed.

\subsection{Hamiltonians for Rydberg Antiblockaded Scar Tower}
\label{apx:rydberg-tower-ham}
The Rydberg antiblockaded tower also has reflection symmetry, so we can carry out a similar calculation as the SU(2) tower.
We carry out the same calculation for the Rydberg scar tower.
For the $(m=0,p=+1)$ sector, we find 4 null vectors:
\begingroup
\allowdisplaybreaks
\begin{align*}
	\hat H_1 &= 2 \hat h_{0,2} + \hat h_{0,1}, \\
	\hat H_2 &= \hat h_{0,6} +\hat h_{0,7}, \\
	\hat H_3 &= \hat h_{0,3}+\hat h_{0,8}, \\
	\hat H_4 &= \hat h_{0,1}-\hat h_{0,3} -\hat h_{0,4}.
\end{align*}
\endgroup
In the sector $(m=0,p=-1)$, we find 2 null vectors:
\begingroup
\allowdisplaybreaks
\begin{align*}
	\hat H_5 &= \hat h_{0,14}, \\
	\hat H_6 &= \hat h_{0,11} +\hat h_{0,12}.
\end{align*}
\endgroup
In the $(m=1,p= +1)$ sector, we find 4 independent solutions:
\begingroup
\allowdisplaybreaks
\begin{align*}
	\hat H_7 &= \hat h_{1,1} - \hat h_{1,3}, \\
	\hat H_8 &= \hat h_{1,2} - \hat h_{1,4}, \\
	\hat H_9 &= 2\hat h_{1,2} + \hat h_{1,6} + \hat h_{1,8} + \hat h_{1,10}, \\
	\hat H_{10} &= 2 \hat h_{1,3} +\hat h_{1,5} +\hat h_{1,7} +\hat h_{1,9}.
\end{align*}
\endgroup
In the sector $(m=1,p=-1)$ sector, we find 2 null vectors:
\begingroup
\allowdisplaybreaks
\begin{align*}
	\hat H_{11} &= \hat h_{1,16} + \hat h_{1,18} +\hat h_{1,20}, \\
	\hat H_{12} &= \hat h_{1,15} + \hat h_{1,17} +\hat h_{1,19}.
\end{align*}
\endgroup
In the $(m=2,p=-1)$ sector, there are 2 null vectors:
\begingroup
\allowdisplaybreaks
\begin{align*}
	\hat H_{13} &= \hat h_{2,1}, \\
	\hat H_{14} &= \hat h_{2,2}.
\end{align*}
\endgroup
And there are no solution in the sectors $(m=2,p=+1)$ and $(m=3)$.
In together, there are 14 linearly independent solutions.

The scar Hamiltonian in Refs.~\cite{domain-wall-Tomasi,domain-wall} for the Rydberg antiblockaded tower is:
\begin{equation}
\begin{aligned}
	\hat H_{\mathrm{Rydberg}} &= \lambda \sum_j (\sigma_j^x-\sigma_{j-1}^z\sigma_j^x \sigma_{j+1}^z) + \\ 
	&\Delta\sum_j \sigma_j^z + J\sum_j \sigma_j^z \sigma_{j+1}^z.
\end{aligned}
\end{equation}
The first term is nothing but the $\hat H_7$, the second and the third term can be re-expressed as:
\begin{equation}
	\frac{J}{2}\hat H_1 + (\Delta-J) \hat H^z,
\end{equation}
where $\hat H^z$ is the generator of the U(1) group as the spectrum-splitting term.
Thus we show this Hamiltonian also falls into the general form of solution.

\subsection{Hamiltonians for Onsager Scar Tower}
\label{apx:onsager-tower-ham}
The Onsager scar tower also has reflection symmetry, which allows us to decompose the covariant matrix into different parity sectors. 
Following the same routine, in the $(m=0,p=+1)$ sector, we find 3 null vectors:
\begingroup
\allowdisplaybreaks
\begin{align*}
	\hat H_{1} &= \hat h_{0,1} + 2\hat h_{0,2} -\hat h_{0,3}- \hat h_{0,4}, \\
	\hat H_{2} &= \hat h_{0,2} - \hat h_{0,6} -\hat h_{0,7}, \\
	\hat H_{3} &= \hat h_{0,5}.
\end{align*}
\endgroup
In the $(m=0,p=-1)$ sector, we find 2 null vectors:
\begingroup
\allowdisplaybreaks
\begin{align*}
	\hat H_{4} &= \hat h_{0,11}, \\
	\hat H_{5} &= \hat h_{0,12}.
\end{align*}
\endgroup
In the $(m=1,p=+1)$ sector, we find 2 null vectors:
\begingroup
\allowdisplaybreaks
\begin{align*}
	\hat H_{6} &= 2\hat h_{1,1}+\hat h_{1,5}-\hat h_{1,7}-\hat h_{1,9}, \\
	\hat H_{7} &= 2\hat h_{1,2} +\hat h_{1,6}-\hat h_{1,8}-\hat h_{1,10}.
\end{align*}
\endgroup
And there is no null vector in the sector $(m=1,p=-1)$, $(m=2,p=\pm1)$ and $(m=3)$.
In together, there are 7 independent solutions.
A general scar Hamiltonian is a vector in this 7-dimensional vector space:
\begin{equation}
	\hat H(\{J_i\}) = \sum_{i=1}^7 J_i \hat H_i.
\end{equation}

In Ref.~\cite{onsager}, the Hamiltonian for such tower is
\begin{widetext}
\begin{equation}
	\hat H_{\mathrm{Onsager}} = \sum_j (\sigma_j^+\sigma_{j+1}^-+\sigma_j^-\sigma_{j+1}^+) + h\sum_j \sigma_j^z + \hat H_{\mathrm{pert}},
\end{equation}
The first term is $\hat H_3$, and the second term is the spectrum-splitting term $\hat H^z$.
We then examine the perturbation term:
\begin{equation}
	\hat H_{\mathrm{pert}} = \sum_j \left\{ c_1\left|\downarrow\uparrow\downarrow\right\rangle\left \langle\downarrow\uparrow\downarrow\right|+
	 c_2 \frac{\left|\downarrow\uparrow\uparrow\right\rangle + \left|\uparrow\uparrow\downarrow\right\rangle}{\sqrt 2} \frac{\left \langle\downarrow\uparrow\uparrow\right| + \left \langle\uparrow\uparrow\downarrow\right|}{\sqrt 2} + 
	c_3 \left(\left|\downarrow\uparrow\downarrow\right\rangle \frac{\left \langle\downarrow\uparrow\uparrow\right| + \left \langle\uparrow\uparrow\downarrow\right|}{\sqrt 2} + h.c.\right) \right\}_{j-1,j,j+1}.
\end{equation}
Note in the Ref.~\cite{onsager}, the coefficients are actually site-dependent, but we only consider the translational invariant case here.
The first term in the perturbation term written as the spin operator is 
\begingroup
\allowdisplaybreaks
\begin{eqnarray}
	\sum_j (\left|\downarrow\uparrow\downarrow\right\rangle\left \langle\downarrow\uparrow\downarrow\right|)_{j-1,j,j+1} 
	&=& \sum_j \frac{1-\sigma_{j-1}^z}{2} \frac{1+\sigma_{j}^z}{2} \frac{1-\sigma_{j+1}^z}{2} \nonumber \\
	&=& \frac{1}{8} \sum_j \left[1 -\sigma_j^z-2\sigma_j^z\sigma_{j+1}^z+\sigma_{j-1}^z\sigma_{j+1}^z+\sigma_{j-1}^z\sigma_j^z\sigma_{j+1}^z \right] \nonumber \\
	&=& -\frac{1}{8} \hat H_1 + const.
\end{eqnarray}
\endgroup
The second term is
\begingroup
\allowdisplaybreaks 
\begin{eqnarray}
	&&\sum_j \left(\frac{\left|\downarrow\uparrow\uparrow\right\rangle + \left|\uparrow\uparrow\downarrow\right\rangle}{\sqrt 2} \frac{\left \langle\downarrow\uparrow\uparrow\right| + \left \langle\uparrow\uparrow\downarrow\right|}{\sqrt 2}\right)_{j-1,j,j+1} \nonumber \\
	&=& \sum_j \frac{1+\sigma_j^z}{2} 
	\left(\frac{\left|\downarrow\uparrow\right\rangle + \left|\uparrow\downarrow\right\rangle}{\sqrt 2} \frac{\left \langle\downarrow\uparrow\right| + \left \langle\uparrow\downarrow\right|}{\sqrt 2}\right)_{j-1,j+1} \nonumber \\
	&=& \frac{1}{2}\sum_j \frac{1+\sigma_j^z}{2} \left(\frac{1-\sigma_{j-1}^z}{2}\frac{1+\sigma_{j+1}^z}{2} + \frac{1+\sigma_{j-1}^z}{2}\frac{1-\sigma_{j+1}^z}{2} +\sigma_{j-1}^+\sigma_{j+1}^- + \sigma_{j-1}^-\sigma_{j+1}^+\right) \nonumber \\
	&=& \frac{\hat h_{0,6} +\hat h_{0,7}}{4} + \frac{1}{8} \sum_j\left(1+\sigma_j^z-\sigma_{j-1}^z\sigma_{j+1}^z-\sigma_{j}^z\sigma_{j-1}^z\sigma_{j+1}^z\right) \nonumber \\
	&=& \frac{\hat h_{0,6} +\hat h_{0,7}}{4} -\frac{\hat H_1 - 2\hat h_{0,2}}{8} +const. \nonumber \\
	&=& \frac{1}{4}\hat H_2 -\frac{1}{8}\hat H_1 +const.
\end{eqnarray}
The third term is
\endgroup
\begingroup
\allowdisplaybreaks
\begin{eqnarray}
		&&\sum_j \left(\left|\downarrow\uparrow\downarrow\right\rangle \frac{\left \langle\downarrow\uparrow\uparrow\right| + \left \langle\uparrow\uparrow\downarrow\right|}{\sqrt 2} + h.c.\right)_{j-1,j,j+1} \nonumber \\
	&=& \sum_j \frac{1+\sigma_j^z}{2} 
	\left(\left|\downarrow\downarrow\right\rangle \frac{\left \langle\downarrow\uparrow\right| + \left \langle\uparrow\downarrow\right|}{\sqrt 2} + \frac{\left|\downarrow\uparrow\right\rangle + \left|\uparrow\downarrow\right\rangle}{\sqrt 2}\left\langle\downarrow\downarrow\right|\right)_{j-1,j+1} \nonumber \\
	&=& \frac{1}{\sqrt 2}\sum_j \frac{1+\sigma_j^z}{2} 
	\left(\frac{1-\sigma_{j-1}^z}{2}\sigma_{j+1}^x + \sigma_{j-1}^x\frac{1-\sigma_{j+1}^z}{2}\right) \nonumber \\
	&=& \frac{1}{8\sqrt 2}\left(2\hat h_{1,1}+\hat h_{1,5}-\hat h_{1,7}-\hat h_{1,9}\right) \nonumber \\
	&=& \frac{1}{8\sqrt 2} \hat H_6.
\end{eqnarray}
\endgroup
\end{widetext}
Thus the Hamiltonian $\hat H_{\mathrm{Onsager}}$ falls into the general solution.

\subsection{Adjoint Representation of U(1) on Two-site Spin-1 Operators} 
For spin-1 system, we can choose the local operator basis as:
\begin{equation}
\begin{aligned}
	\alpha &\equiv \left[\begin{array}{ccc}
		1 & 0 & 0 \\
		0 & -1 & 0 \\
		0 & 0 & 0
	\end{array}\right], \
	\beta \equiv \left[\begin{array}{ccc}
		0 & 0 & 0 \\
		0 & 1 & 0 \\
		0 & 0 & -1
	\end{array}\right], \\
	u &\equiv |+\rangle\langle0|, \
	v \equiv |0\rangle\langle-|, \
	w \equiv |+\rangle\langle-|, \\
	\bar u &\equiv |0\rangle \langle +|,\ 
	\bar v \equiv |-\rangle \langle0|,\ 
	\bar w\equiv |-\rangle\langle +|.
\end{aligned}
\end{equation}
Each operator gives a one-dimensional adjoint representation of U(1):
\begin{equation}
\begin{aligned}
	e^{i\theta\hat S^z} \alpha e^{-i\theta\hat S^z} &= \alpha, & 
	e^{i\theta\hat S^z} \beta e^{-i\theta\hat S^z} &= \beta,\\
	e^{i\theta\hat S^z} u e^{-i\theta\hat S^z} &= e^{i\theta}u, & 
	e^{i\theta\hat S^z} \bar u e^{-i\theta\hat S^z} &= e^{-i\theta} \bar u, \\
	e^{i\theta\hat S^z} v e^{-i\theta\hat S^z} &= e^{i\theta}v, & 
	e^{i\theta\hat S^z} \bar v e^{-i\theta\hat S^z} &= e^{-i\theta}\bar v, \\
	e^{i\theta\hat S^z} w e^{-i\theta\hat S^z} &= e^{2i\theta}w, & 
	e^{i\theta\hat S^z} \bar w e^{-i\theta\hat S^z} &= e^{-2i\theta}\bar w.
\end{aligned}
\end{equation}
On one-site (traceless) operator space, there are one $(2\oplus -2)$ representation, two $(1\oplus -1)$ representation, and two $0$ representation.
On two-site (traceless) operator space, the Clebsch-Gordan decomposition gives:
\begin{equation}
\begin{aligned}
	&[2\oplus (1\times 2) \oplus (0\times 2) \oplus (-1\times 2)\oplus -2]^{\otimes 2} \\
	=& [4\oplus-4] \oplus [(3\oplus-3)\times 4] \oplus [(2\oplus-2)\times 8] \oplus \\
	 & [(1\oplus-1)\times 12] \oplus [0\times 14].
\end{aligned}
\end{equation}
In the $0$ representation, there are $(14+2=16)$ independent Hermitian operators, including 10 reflection-symmetric terms:
\begingroup
\allowdisplaybreaks
\begin{align*}
	\hat h_{0,1}^{[j]} &= \alpha_j, \\
	\hat h_{0,2}^{[j]} &= \beta_j, \\
	\hat h_{0,3}^{[j]} &= \alpha_j \alpha_{j+1}, \\
	\hat h_{0,4}^{[j]} &= \beta_j \beta_{j+1}, \\
	\hat h_{0,5}^{[j]} &= \alpha_j \beta_{j+1} + \beta_j \alpha_{j+1}, \\
	\hat h_{0,6}^{[j]} &= u_j \bar u_{j+1} + \bar u_j u_{j+1}, \\
	\hat h_{0,7}^{[j]} &= v_j \bar v_{j+1} + \bar v_j v_{j+1}, \\
	\hat h_{0,8}^{[j]} &= u_j \bar v_{j+1} + \bar v_j u_{j+1} +h.c., \\
	\hat h_{0,9}^{[j]} &= \frac{u_j \bar v_{j+1}+\bar v_j u_{j+1}}{i} + h.c., \\
	\hat h_{0,10}^{[j]} &= w_j \bar w_{j+1} + \bar w_j w_{j+1}, 
\end{align*}
\endgroup
and 6 antisymmetry terms:
\begingroup
\allowdisplaybreaks
\begin{align*}
	\hat h_{0,11}^{[j]} &= \alpha_j\beta_{j+1}-\beta_j\alpha_{j+1}, \\
	\hat h_{0,12}^{[j]} &= \frac{u_j \bar u_{j+1} - \bar u_j u_{j+1}}{i}, \\
	\hat h_{0,13}^{[j]} &= \frac{v_j \bar v_{j+1} - \bar v_j v_{j+1}}{i}, \\
	\hat h_{0,14}^{[j]} &= u_j \bar v_{j+1} - \bar v_j u_{j+1} +h.c., \\
	\hat h_{0,15}^{[j]} &= \frac{u_j \bar v_{j+1} - \bar v_j u_{j+1}}{i} + h.c., \\
	\hat h_{0,16}^{[j]} &= w_j \bar w_{j+1} - \bar w_j w_{j+1}.
\end{align*}
\endgroup
In the $(1\oplus -1)$ representation, there are $(12 \times 2+2 \times 2 = 28)$ independent Hermitian operators, including 16 reflection-symmetric terms:
\begingroup
\allowdisplaybreaks
\begin{align*}
	\hat h_{1,1}^{[j]} &= u_j + \bar u_j, \\
	\hat h_{1,2}^{[j]} &= \frac{u_j - \bar u_j}{i}, \\
	\hat h_{1,3}^{[j]} &= v_j + \bar v_j, \\
	\hat h_{1,4}^{[j]} &= \frac{v_j - \bar v_j}{i}, \\
	\hat h_{1,5}^{[j]} &= u_j \alpha_{j+1} + \alpha_j u_{j+1} + h.c., \\
	\hat h_{1,6}^{[j]} &= \frac{u_j \alpha_{j+1} + \alpha_j u_{j+1}}{i} + h.c., \\
	\hat h_{1,7}^{[j]} &= u_j\beta_{j+1} + \beta_j u_{j+1} + h.c., \\
	\hat h_{1,8}^{[j]} &= \frac{u_j\beta_{j+1} + \beta_j u_{j+1}}{i}+ h.c., \\
	\hat h_{1,9}^{[j]} &= v_j\alpha_{j+1} + \alpha_j v_{j+1} + h.c., \\
	\hat h_{1,10}^{[j]} &= \frac{v_j\alpha_{j+1} + \alpha_j v_{j+1}}{i} +h.c., \\
	\hat h_{1,11}^{[j]} &= v_j\beta_{j+1} + \beta_jv_{j+1} + h.c., \\
	\hat h_{1,12}^{[j]} &= \frac{v_j\beta_{j+1} + \beta_jv_{j+1}}{i} +h.c., \\
	\hat h_{1,13}^{[j]} &= w_j \bar u_{j+1} + \bar u_j w_{j+1} + h.c., \\
	\hat h_{1,14}^{[j]} &= \frac{w_j \bar u_{j+1} + \bar u_j w_{j+1}}{i} +  h.c., \\
	\hat h_{1,15}^{[j]} &= w_j \bar v_{j+1} + \bar v_j w_{j+1} + h.c., \\
	\hat h_{1,16}^{[j]} &= \frac{w_j \bar v_{j+1} + \bar v_j w_{j+1}}{i} +  h.c.,
\end{align*}
\endgroup
and 12 antisymmetric terms:
\begingroup
\allowdisplaybreaks
\begin{align*}
	\hat h_{1,17}^{[j]} &= u_j \alpha_{j+1} - \alpha_j u_{j+1} + h.c., \\
	\hat h_{1,18}^{[j]} &= \frac{u_j \alpha_{j+1} - \alpha_j u_{j+1}}{i} + h.c., \\
	\hat h_{1,19}^{[j]} &= u_j\beta_{j+1} - \beta_j u_{j+1} + h.c., \\
	\hat h_{1,20}^{[j]} &= \frac{u_j\beta_{j+1} - \beta_j u_{j+1}}{i}+ h.c., \\
	\hat h_{1,21}^{[j]} &= v_j\alpha_{j+1} - \alpha_j v_{j+1} + h.c., \\
	\hat h_{1,22}^{[j]} &= \frac{v_j\alpha_{j+1} - \alpha_j v_{j+1}}{i} +h.c., \\
	\hat h_{1,23}^{[j]} &= v_j\beta_{j+1} - \beta_jv_{j+1} + h.c., \\
	\hat h_{1,24}^{[j]} &= \frac{v_j\beta_{j+1} - \beta_jv_{j+1}}{i} +h.c., \\
	\hat h_{1,25}^{[j]} &= w_j \bar u_{j+1} - \bar u_j w_{j+1} + h.c., \\
	\hat h_{1,26}^{[j]} &= \frac{w_j \bar u_{j+1} - \bar u_j w_{j+1}}{i} +  h.c., \\
	\hat h_{1,27}^{[j]} &= w_j \bar v_{j+1} - \bar v_j w_{j+1} + h.c., \\
	\hat h_{1,28}^{[j]} &= \frac{w_j \bar v_{j+1} - \bar v_j w_{j+1}}{i} +  h.c.,
\end{align*}
\endgroup
In the $(2\oplus -2)$ representation, there are $(8\times 2 + 1\times 2 = 18)$ independent Hermitian operators, including 12 reflection-symmetric terms:
\begingroup
\allowdisplaybreaks
\begin{align*}
	\hat h_{2,1}^{[j]} &= w_j + \bar w_j, \\
	\hat h_{2,2}^{[j]} &= \frac{w_j - \bar w_j}{i}, \\
	\hat h_{2,3}^{[j]} &= w_j \alpha_{j+1} + \alpha_j w_{j+1} + h.c., \\
	\hat h_{2,4}^{[j]} &= \frac{w_j \alpha_{j+1} + \alpha_j w_{j+1}}{i} +h.c., \\
	\hat h_{2,5}^{[j]} &= w_j \beta_{j+1} + \beta_j w_{j+1} + h.c., \\
	\hat h_{2,6}^{[j]} &= \frac{w_j \beta_{j+1} + \beta_j w_{j+1}}{i} +h.c., \\
	\hat h_{2,7}^{[j]} &= u_j u_{j+1} + \bar u_j \bar u_{j+1}, \\
	\hat h_{2,8}^{[j]} &= \frac{u_j u_{j+1} - \bar u_j \bar u_{j+1}}{i}, \\
	\hat h_{2,9}^{[j]} &= v_j v_{j+1} + \bar v_j \bar v_{j+1}, \\
	\hat h_{2,10}^{[j]} &= \frac{v_j v_{j+1} - \bar v_j \bar v_{j+1}}{i}, \\
	\hat h_{2,11}^{[j]} &= u_j v_{j+1} + v_j u_{j+1} + h.c., \\
	\hat h_{2,12}^{[j]} &= \frac{u_j v_{j+1} +  v_j u_{j+1}}{i} +h.h., 
\end{align*}
\endgroup
and 6 antisymmetric terms:
\begingroup
\allowdisplaybreaks
\begin{align*}
	\hat h_{2,13}^{[j]} &= w_j \alpha_{j+1} - \alpha_j w_{j+1} + h.c., \\
	\hat h_{2,14}^{[j]} &= \frac{w_j \alpha_{j+1} - \alpha_j w_{j+1}}{i} +h.c., \\
	\hat h_{2,15}^{[j]} &= w_j \beta_{j+1} - \beta_j w_{j+1} + h.c., \\
	\hat h_{2,16}^{[j]} &= \frac{w_j \beta_{j+1} - \beta_j w_{j+1}}{i} +h.c., \\
	\hat h_{2,17}^{[j]} &= u_j v_{j+1} - v_j u_{j+1} + h.c., \\
	\hat h_{2,18}^{[j]} &= \frac{u_j v_{j+1} -  v_j u_{j+1}}{i} +h.h., 
\end{align*}
\endgroup
In the $(3 \oplus -3)$ representation, there are $(4 \times 2 = 8)$ independent Hermitian terms, including 4 reflection-symmetric terms:
\begingroup
\allowdisplaybreaks
\begin{align*}
	\hat h_{3,1}^{[j]} &= w_j u_{j+1} + u_j w_{j+1} + h.c., \\
	\hat h_{3,2}^{[j]} &= \frac{w_j u_{j+1} + u_j w_{j+1}}{i} +h.c., \\
	\hat h_{3,3}^{[j]} &= w_j v_{j+1} + v_j w_{j+1} + h.c., \\
	\hat h_{3,4}^{[j]} &= \frac{w_j v_{j+1} + v_j w_{j+1}}{i} +h.c.,
\end{align*}
\endgroup
and 4 antisymmetric terms:
\begingroup
\allowdisplaybreaks
\begin{align*}
	\hat h_{3,5}^{[j]} &= w_j u_{j+1} - u_j w_{j+1} + h.c., \\
	\hat h_{3,6}^{[j]} &= \frac{w_j u_{j+1} - u_j w_{j+1}}{i} +h.c., \\
	\hat h_{3,7}^{[j]} &= w_j v_{j+1} - v_j w_{j+1} + h.c., \\
	\hat h_{3,8}^{[j]} &= \frac{w_j v_{j+1} - v_j w_{j+1}}{i} +h.c.,
\end{align*}
\endgroup
In the $(4\oplus-4)$ representation, there are $(1\times 2=2)$ reflection-symmetric terms:
\begingroup
\allowdisplaybreaks
\begin{align*}
	\hat h_{4,1}^{[j]} &= w_j w_{j+1} + \bar w_j \bar w_{j+1}, \\
	\hat h_{4,2}^{[j]} &= \frac{w_j w_{j+1} - \bar w_j \bar w_{j+1}}{i}.
\end{align*}
\endgroup

\subsection{Hamiltonians for the Scar Tower of the Spin-1 XY Model}
\label{apx:xy-1-tower-ham}
The scar tower of the spin-1 XY model is reflection symmetric, and thus we can do a similar calculation as in the spin-1/2 cases.
The covariant matrix is computed on the $(L=8)$ system.
In the sector $(m=0,p=+1)$, we find 8 independent solutions:
\begingroup
\allowdisplaybreaks
\begin{align*}
	\hat H_1 &= \hat h_{0,6}, \\
	\hat H_2 &= \hat h_{0,7}, \\
	\hat H_3 &= \hat h_{0,8}, \\
	\hat H_4 &= \hat h_{0,9}, \\
	\hat H_5 &= \hat h_{0,1} - \hat h_{0,2}, \\
	\hat H_6 &= \hat h_{0,5} - \hat h_{0,10}, \\
	\hat H_7 &= \hat h_{0,3} + \hat h_{0,4} - \hat h_{0,5}, \\
	\hat H_8 &= 2\hat h_{0,2}+2\hat h_{0,4}-\hat h_{0,5}.
\end{align*}
\endgroup
In the sector of $(m=0,p=-1)$, there are 4 solutions:
\begingroup
\allowdisplaybreaks
\begin{align*}
	\hat H_9 &= \hat h_{0,11}, \\
	\hat H_{10} &= \hat h_{0,12}, \\
	\hat H_{11} &= \hat h_{0,13}, \\
	\hat H_{12} &= \hat h_{0,16}.
\end{align*}
\endgroup
In the sector $(m=1,p=+1)$, there are 8 solutions:
\begingroup
\allowdisplaybreaks
\begin{align*}
	\hat H_{13} &= 2\hat h_{1,1}-\hat h_{1,5}+\hat h_{1,15}, \\
	\hat H_{14} &= 2\hat h_{1,2}-\hat h_{1,6}+\hat h_{1,16}, \\
	\hat H_{15} &= 2\hat h_{1,3}+\hat h_{1,11}+\hat h_{1,13}, \\
	\hat H_{16} &= 2\hat h_{1,4}+\hat h_{1,12}+\hat h_{1,14}, \\
	\hat H_{17} &= \hat h_{1,7} - \hat h_{1,15}, \\
	\hat H_{18} &= \hat h_{1,8} - \hat h_{1,16}, \\
	\hat H_{19} &= \hat h_{1,9} + \hat h_{1,13}, \\
	\hat H_{20} &= \hat h_{1,10}+\hat h_{1,14}.
\end{align*}
\endgroup
In the sector $(m=1,p=-1)$, there are 8 solutions :
\begingroup
\allowdisplaybreaks
\begin{align*}
	\hat H_{21} &= \hat h_{1,17} + \hat h_{1,27}, \\
	\hat H_{22} &= \hat h_{1,18} - \hat h_{1,20}, \\
	\hat H_{23} &= \hat h_{1,19} + \hat h_{1,27}, \\
	\hat H_{24} &= \hat h_{1,20} + \hat h_{1,28}, \\
	\hat H_{25} &= \hat h_{1,21} - \hat h_{1,25}, \\
	\hat H_{26} &= \hat h_{1,22} - \hat h_{1,26}, \\
	\hat H_{27} &= \hat h_{1,23} - \hat h_{1,25}, \\
	\hat H_{28} &= \hat h_{1,24} - \hat h_{1,26}.
\end{align*}
\endgroup
In the sector $(m=2,p=+1)$, there are 4 solutions:
\begingroup
\allowdisplaybreaks
\begin{align}
	\hat H_{29} &= \hat h_{2,11}, \\
	\hat H_{30} &= \hat h_{2,12}, \\
	\hat H_{31} &= 2\hat h_{2,1} - \hat h_{2,3} + \hat h_{2,5}, \\
	\hat H_{32} &= 2\hat h_{2,2} - \hat h_{2,4} + \hat h_{2,6}.
\end{align}
\endgroup
In the sector $(m=2,p=-1)$, there are 4 solutions:
\begingroup
\allowdisplaybreaks
\begin{align}
	\hat H_{33} &= \hat h_{2,17}, \\
	\hat H_{34} &= \hat h_{2,18}, \\
	\hat H_{35} &= \hat h_{2,13} - \hat h_{2,15}, \\
	\hat H_{36} &= \hat h_{2,14} - \hat h_{2,16}.
\end{align}
\endgroup
And there is no solution in sectors $(m=3)$ and $(m=4)$.
Together there are 36 linearly independent solutions.
The general Hamiltonian is
\begin{equation}
	\hat H(\{J_i\}) = \sum_{i=1}^{36} J_i \hat H_i.
\end{equation}

The Hamiltonian of spin-1 XY model is~\cite{XY-1}:
\begin{equation}
	\hat H_{XY} = \sum_j (\hat S_j^x\hat S_{j+1}^x+\hat S_j^y\hat S_{j+1}^y) + 
	 D \sum_j (\hat S_j^z)^2 + h \sum_j \hat S_j^z.
\end{equation}
The first term is
\begin{eqnarray}
	&& \sum_j (\hat S_j^x\hat S_{j+1}^x+\hat S_j^y\hat S_{j+1}^y) \nonumber \\
	&=& \sum_j (u_j+v_j)(\bar u_{j+1}+\bar v_{j+1}) +h.c. \nonumber \\
	&=& \hat H_1 + \hat H_2 +\hat H_3.
\end{eqnarray}
The second term is
\begin{equation}
	\sum_j (\hat S_j^z)^2 = \frac{1}{3} \hat H_5 + \frac{2}{3}.
\end{equation}
The third term is the spectrum splitting term. 
Thus $\hat H_{\mathrm{XY}}$ is in the general solution.

\subsection{Hamiltonians for the Additional Scar Tower of Spin-1 XY Model}
\label{apx:xy-2-tower-ham}
Similarly, we carry out the calculation on $(L=8)$ system. 
As the result, in the $(m=0,p=+1)$ sector, there are 3 independent solutions:
\begingroup
\allowdisplaybreaks
\begin{align*}
	\hat H_1 &= \hat h_{0,10}, \\
	\hat H_2 &= \hat h_{0,6} + \hat h_{0,7} + \hat h_{0,8}, \\
	\hat H_3 &= 2\hat h_{0,1}-2\hat h_{0,2}-4\hat h_{0,3}-4\hat h_{0,4}-5\hat h_{0,5}.
\end{align*}
\endgroup
In the $(m=0,p=-1)$ sector, there are 2 solutions:
\begingroup
\allowdisplaybreaks
\begin{align*}
	\hat H_4 &= \hat h_{0,11}, \\
	\hat H_5 &= \hat h_{0,16}.
\end{align*}
\endgroup
And there is no more solution in other sectors.
Together, a general scar Hamiltonian is a vector in the 5-dimensional space:
\begin{equation}
	\hat H(\{J_i\}) = \sum_{i=1}^5 J_i \hat H_i.
\end{equation}

The scar Hamiltonian in Ref.~\cite{XY-2} is
\begin{equation}
\begin{aligned}
	\hat H_{XY} &= \sum_j (\hat S_j^x\hat S_{j+1}^x+\hat S_j^y\hat S_{j+1}^y) + \\
	& \epsilon \sum_j \left[(\hat S_j^+)^2(S_{j+1}^-)^2+h.c.\right] + h \sum_j \hat S_j^z.
\end{aligned}
\end{equation}
The first term is $\hat H_2$, the second term is $\hat H_1$, and the third term is the spectrum splitting term.
We thus show $\hat H_{\mathrm{XY}}$ is in the general solutions.

\subsection{Hamiltonians for AKLT Scar Tower}
\label{apx:aklt-tower-ham}
We carry out the same calculation. 
In the sector $(m=0,p=+1)$, there are 4 null vectors:
\begingroup
\allowdisplaybreaks
\begin{align*}
	\hat H_1 &= \hat h_{0,6} - \hat h_{0,7}, \\
	\hat H_2 &= 2\hat h_{0,1}-\hat h_{0,3} + 4\hat h_{0,4}-6\hat h_{0,7}, \\
	\hat H_3 &= 2\hat h_{0,2}-4\hat h_{0,4}-\hat h_{0,5}+3\hat h_{0,7}, \\
	\hat H_4 &= \hat h_{0,3} -8\hat h_{0,4}+2\hat h_{0,5}+9\hat h_{0,7}- 6\hat h_{0,8}-3\hat h_{0,10}.
\end{align*}
\endgroup
In the sector $(m=1,p=+1)$, there are 4 solutions:
\begingroup
\allowdisplaybreaks
\begin{align*}
	\hat H_5 &= 2\hat h_{1,1}-\hat h_{1,5}-2\hat h_{1,7}+6\hat h_{1,11}+3\hat h_{1,15}, \\
	\hat H_6 &= 2\hat h_{1,2}-\hat h_{1,6}-2\hat h_{1,8}+6\hat h_{1,12}+3\hat h_{1,16}, \\
	\hat H_7 &= 2\hat h_{1,3}-\hat h_{1,9}-2\hat h_{1,11}, \\
	\hat H_8 &= 2\hat h_{1,4}-\hat h_{1,10}-2\hat h_{1,12}.
\end{align*}
\endgroup
In the sector $(m=2, p=+1)$, there are 2 solutions:
\begingroup
\allowdisplaybreaks
\begin{align*}
	\hat H_9    &= 2\hat h_{2,1}-\hat h_{2,3}-2\hat h_{2,5}+3\hat h_{2,9}, \\
	\hat H_{10} &= 2\hat h_{2,2}-\hat h_{2,4}-2\hat h_{2,6}+3\hat h_{2,10}.
\end{align*}
\endgroup
And there is no more solution in other sectors.
Together, the general Hamiltonian form an 10-dimensional vector space.

The AKLT Hamiltonian is~\cite{AKLT1987}:
\begin{equation}
	\hat H_{\mathrm{AKLT}} = \sum_j \left[\frac{1}{2}\vec S_j\cdot \vec S_{j+1} +\frac{1}{6}(\vec S_j\cdot \vec S_{j+1})^2+\frac{1}{3} \right],
\end{equation}
which can be re-expressed as
\begin{equation}
	\hat H_{\mathrm{AKLT}} = \frac{\hat H_1 -\hat H_2 - \hat H_3}{2} + \frac{\hat H_4}{9} + \sum_j \hat S^z_j.
\end{equation}
We thus showed $\hat H_{\mathrm{AKLT}}$ falls into the general solution.

\section{Ladder Operators for the Scar Towers}
\label{apx:ladder-operator}
Several known scar models have ladder operators that generate the tower states.
In our deforming framework, we regard those towers as deformed SU(2)-symmetric spaces.
Such structure does not necessarily have a ladder operator, while since it gives a clearer picture of the scar space and helps compare the towers we construct with those existing ones, we in the following give a sufficient condition for the existence of ladder operator based on the MPS representation of tower states. We then use the condition to explicitly reveal the ladder operators for the towers constructed from our framework.

\subsection{Sufficient Condition for the Existence of Ladder Operator}
\label{apx:tower-condition}
Here we consider the simplest case where the towers are translational invariant, and the local HWS gives the 2-dimensional representation of $\mathfrak{su}(2)$.
The MPS tensor of the deformed highest-weight state and the excited tensor are denoted as $A$ and $B$.
A general tower state is
\begin{equation}
	|\Psi_n\rangle = \sum_{\{X^{[i]}\}}C_{\{X^{[i]}\}}\sum_{\{s_i\}}\mathrm{Tr}\left[\prod_{i=1}^N X^{[i]}_{s_i} \right]|\{s_i\}\rangle,
\end{equation}
where among $N$ tensors $\{X^{[i]}\}$ there are $n$ excited ones (assumed to locate at sites $\{e_1, e_2, \dots, e_n\}$.
The tower is the superposition of all possible configurations, with the coefficients
\begin{equation}
	C_{\{X^{[i]}\}} = \exp\left(i k \sum_{j=1}^n e_j\right).
\end{equation}
Consider a contiguous cluster with ($m+n+1$) sites with an excited tensor in the middle.
Define two blocked tensors $M$, $\tilde M$ as:
\begin{equation}
\begin{aligned}
	M^{[-m,\dots,n]}_{i_{-m},\dots,i_n}
	& \equiv X^{[-m]}_{i_{-m}}\cdots X^{[-1]}_{i_{-1}}A^{[0]}_{i_0}X^{[1]}_{i_1}\cdots X^{[n]}_{i_n}, \\
	\tilde M^{[-m,\dots,n]}_{i_{-m},\dots,i_n}
	& \equiv X^{[-m]}_{i_{-m}}\cdots X^{[-1]}_{i_{-1}}B^{[0]}_{i_0}X^{[1]}_{i_1}\cdots X^{[n]}_{i_n},
\end{aligned}
\end{equation}
where all other tensor $X^{[i]}$ can be $A$ or $B$.
If the blocked tensor $M$ can be generated by a $(m+n+1)$-site local operator, regardless of other tensor $X^{[i]}$, i.e.,
\begin{equation}
\begin{aligned}
	\tilde M^{[-m,\dots,n]}_{i_{-m},\dots,i_n} 
	&= \sum_{\{j_k\}} \left(\hat q^+\right)_{i_{-m},\dots,i_n}^{j_{-m},\dots,j_n} 
	M^{[-m,\dots,n]}_{j_{-m},\dots,j_n} \label{eq:local-excitation-condition}
\end{aligned}
\end{equation}
for all $X^{[i]}$ configurations. 
When such condition is satisfied, the tensor excitation can be lifted to a physical excitation, so that
\begin{equation}
	|\Psi_n\rangle = \sum_{\{e_j\}}e^{ik \sum_j e_j} \left[\bigotimes_{j=1}^n \hat q^+_{e_j}\right] |\Psi_0\rangle \label{eq:generating-tower}
\end{equation}
The local excitations $\hat q^+_j$ do not necessarily commute with each other.
However, the condition (\ref{eq:local-excitation-condition}) ensures the commutation of $\hat q^+_j$'s when restricted to the tower space.
In this way, Eq.~(\ref{eq:generating-tower}) can be expressed in a compact form:
\begin{equation}
	|\Psi_n\rangle = \left(\sum_j e^{ikj} \hat q^+_j\right)^n|\Psi_0\rangle
\end{equation}
The ladder operator is
\begin{equation}
	\hat Q^+ \equiv \sum_j e^{ikj} \hat q^+_j.
\end{equation}

\subsection{Ladder Operator for Type-2 Tower}
\label{apx:type-2-ladder}
For the special choice of parameters (\ref{eq:rydberg-coeff}), the deforming MPO becomes:
\begin{equation}
	W = \left[\begin{array}{cc}
		\left|\uparrow\rangle \langle \uparrow \right| & \hat\sigma^+ \\
		\hat\sigma^- & \left|\downarrow\rangle \langle \downarrow \right|
	\end{array}\right].
	\label{eq:Rydberg-W}
\end{equation}
The tensor $A$, $B$ are:
\begin{equation}
	\begin{aligned}
		A_{\uparrow} &= \sigma^+, & 
		A_{\downarrow} &= \frac{1-\sigma^z}{2}, \\
		B_{\uparrow} &= 0, & 
		B_{\downarrow} &= \sigma^-.
	\end{aligned}
\end{equation}
It follows immediately from the MPS tensor that the deformed anchor state is the fully polarized state:
\begin{equation}
	|\Psi_0\rangle = \left|\downarrow \cdots \downarrow\right\rangle.
\end{equation}
To investigate the structure of the tower, we consider two three-site clusters:
\begin{equation}
\begin{aligned}	
	M^{[0,1,2]} &\equiv A^{[0]}_{i_0} X^{[1]}_{i_1} X^{[2]}_{i_2}, \\
	\tilde M^{[0,1,2]} &\equiv B^{[0]}_{i_0} X^{[1]}_{i_1} X^{[2]}_{i_2}.
\end{aligned}
\end{equation}
When $X^{[1]}=X^{[2]}=A$ the nonzero elements are:
\begin{align}
	M_{222} = \tilde M_{212} = \frac{1-\sigma^z}{2}.
\end{align}
When $X^{[1]}=A$, $X^{[2]}=B$, the nonzero elements are:
\begin{align}
	M_{222} = \tilde M_{212} = \sigma^-.
\end{align}
When $X^{[1]}=B$, $X^{[2]}=A$, the only nonzero element is:
\begin{align}
	M_{221} = \frac{1-\sigma^z}{2}.
\end{align}
When $X^{[1]}=B$, $X^{[2]}=B$, there is no nonzero element.
In all cases, the condition (\ref{eq:local-excitation-condition}) hold for
\begin{equation}
	\hat q^+_j = \hat{P}^{\downarrow}_{j-1} \sigma_j^+ \hat{P}^{\downarrow}_{j+1}. 
\end{equation}
In this way, the tower of states has a ladder operator:
\begin{equation}
	\hat Q^+_{\mathrm{Rydberg}} = \sum_j (-1)^j \hat{P}^{\downarrow}_{j-1} \sigma_j^+ \hat{P}^{\downarrow}_{j+1},
\end{equation}
which is exactly the ladder operator of the scar tower in Ref.~\cite{domain-wall}.

\subsection{Ladder Operator for Type-3 Tower}
\label{apx:type-3-ladder}
For the choice of parameter (\ref{eq:onsager-coeff}), the deforming MPO is
\begin{equation}
	W = \left[\begin{array}{cc}
		\left|\downarrow\rangle \langle \downarrow \right| &  \sigma^+  \\
		\left|\uparrow\rangle \langle \uparrow \right| & 0 
	\end{array}\right].
\end{equation}
The corresponding MPS tensors $A$, $B$ for the parameters (\ref{eq:onsager-coeff}) are:
\begin{equation}
	\begin{aligned}
		A_{\uparrow} &= \sigma^+, &
		A_{\downarrow} &= \frac{1+\sigma^z}{2}, \\
		B_{\uparrow} &= \sigma^-, &
		B_{\downarrow} &= 0.
	\end{aligned}
\end{equation}
The deformed anchor is again
\begin{equation}
	|\Psi_{\bm M}\rangle = \left|\downarrow \cdots \downarrow\right\rangle.
\end{equation}
We then consider two two-site clusters:
\begin{equation}
\begin{aligned}
	M^{[-1,0]} &\equiv X^{[-1]}_{i_0} A^{[0]}_{i_1}, \\
	\tilde M^{[-1,0]} &\equiv X^{[-1]}_{i_0} B^{[0]}_{i_1}.
\end{aligned}
\end{equation}
When $X^{[-1]} = A$, the nonzero elements are:
\begin{equation}
	M_{21} = \sigma^+,\ M_{22}=\tilde M_{11}=\frac{1+\sigma^z}{2}
\end{equation}
When $X^{[-1] = B}$, the nonzero elements are:
\begin{equation}
	M_{12} = \sigma^-,\ M_{11}=\frac{1+\sigma^z}{2}.
\end{equation}
In both case (\ref{eq:local-excitation-condition}) hold for 
where the two-site operator
\begin{equation}
	\hat q^+_j = \hat \sigma_j^+ \hat \sigma_{j+1}^+.
\end{equation}
We thus show this tower has a ladder operator:
\begin{equation}
	\hat Q^+_{\mathrm{Onsager}} = \sum_j (-1)^j \hat \sigma_j^+ \hat \sigma_{j+1}^+,
\end{equation}
which is identical to that in Ref.~\cite{onsager}.

\subsection{Ladder Operator for Type-6 Tower}
\label{apx:type-6-ladder}
The deforming MPO tensor under coefficient (\ref{eq:aklt-coeff}) is
\begin{equation}
	W = \left[\begin{array}{cc}
		-\frac{1}{\sqrt 2}|0\rangle\langle-| & 1|+\rangle\langle-| \\
		-|+\rangle\langle+|-|-\rangle\langle-| & \frac{1}{\sqrt 2}|0\rangle\langle -|
	\end{array}\right].
\end{equation}
The matrix-product anchor has the MPS tensor (after renormalization):
\begin{equation}
	A_{\pm} = \pm \sqrt{\frac{2}{3}} \sigma^\pm,\
	A_{0} = -\sqrt{\frac{1}{3}} \sigma^z,
\end{equation}
which is exactly the MPS tensor for AKLT ground state.
Besides, the excited tensor $B$ is
\begin{equation}
	B_{+} = - \sqrt{\frac{2}{3}} \sigma^-,\ 
	B_{0} = B^{[-]} = 0,
\end{equation}
which can be locally generated from tensor $A$ by
\begin{equation}
	\hat q_j^+ = (\hat S_j^+)^2.
\end{equation}
For the prototype tower with $\pi$-momentum excitation, the ladder operator is
\begin{equation}
	\hat Q_{\mathrm{AKLT}}^+ = \sum_j (-1)^j (\hat S_j^+)^2.
\end{equation}
This is exactly the ladder operator of the AKLT scar tower~\cite{AKLT-1,AKLT-2}, and thus we have proved that the deformed tower here is identical to the AKLT scar tower. 

Similarly, for the generalized coefficient (\ref{eq:gen-aklt-coeff}), 
\begin{equation}
	W = \left[\begin{array}{cc}
		c_0|0\rangle\langle-| & c_+ |+\rangle\langle-| \\
		c_- |+\rangle\langle+|+c_-|-\rangle\langle-| & -c_0 |0\rangle\langle -|
	\end{array}\right].
\end{equation}
The anchor state has the MPS tensor:
\begin{equation}
	A_{\pm} = c_{\pm} \sigma^\pm,\
	A_{0} = c_0 \sigma^z,
\end{equation}
and the excited tensor $B$ only has a nonzero element
\begin{equation}
	B^{[+]} = c_- \sigma^-,
\end{equation}
which is also locally generated by $\hat q^+_j$.
We have thus shown that this deformed tower is the generalized AKLT scar tower in Ref.~\cite{SGA-AKLT}.

\end{appendix}
\bibliography{ref}

\end{document}